\documentclass[twocolumn]{aastex631}

\usepackage{graphicx}
\usepackage{amsmath}
\usepackage{float}
\usepackage{natbib}
\usepackage{tabularx}
\usepackage{bm}
\usepackage{appendix}
\usepackage{longtable}
\usepackage{url}

\newcommand\msun{M_{\odot}}
\newcommand\cmc{\texttt{CMC}}

\shorttitle{MSPs in Globular Clusters and the Galactic Center}
\shortauthors{Ye \& Fragione}

\graphicspath{{./}{figures/}}

\begin{document}

\title{Millisecond Pulsars in Dense Star Clusters: Evolution, Scaling Relations, and the Galactic-Center Gamma-ray Excess}

\author[0000-0001-9582-881X]{Claire S. Ye}
\affil{Department of Physics \& Astronomy, Northwestern University, Evanston, IL 60208, USA}
\affil{Center for Interdisciplinary Exploration \& Research in Astrophysics (CIERA), Northwestern University, Evanston, IL 60208, USA}
\affil{Canadian Institute for Theoretical Astrophysics, University of Toronto, 60 St. George Street, Toronto, Ontario M5S 3H8, Canada}
\correspondingauthor{Claire S.~Ye}
\email{shiye2015@u.northwestern.edu}

\author[0000-0002-7330-027X]{Giacomo Fragione}
\affil{Department of Physics \& Astronomy, Northwestern University, Evanston, IL 60208, USA}
\affil{Center for Interdisciplinary Exploration \& Research in Astrophysics (CIERA), Northwestern University, Evanston, IL 60208, USA}

\begin{abstract}
The number of millisecond pulsars (MSPs) observed in Milky Way globular clusters has increased explosively in recent years, but the underlying population is still uncertain due to observational biases. We use state-of-the-art $N$-body simulations to study the evolution of MSP populations in dense star clusters. These cluster models span a wide range in initial conditions, including different initial masses, metallicities, and virial radii, which nearly cover the full range of properties exhibited by the population of globular clusters in the Milky Way. We demonstrate how different initial cluster properties affect the number of MSPs, for which we provide scaling relations as a function of cluster age and mass. As an application, we use our formulae to estimate the number of MSPs delivered to the Galactic Center from inspiralling globular clusters to probe the origin of the Galactic-Center gamma-ray excess detected by \textit{Fermi}. We predict about $400$ MSPs in the Galactic Center from disrupted globular clusters, which can potentially explain most of the observed gamma-ray excess.
\end{abstract}

\section{Introduction} \label{sec:intro}
Since the discovery of a highly-magnetized, fast-spinning, radio-pulsating neutron star (NS) around $50$ years ago \citep{Hewish+1968}, our understanding of these stellar remnants has grown tremendously. They can now be observed across the electromagnetic spectrum, from the radio to X-rays and gamma-rays, and even in gravitational waves \citep[e.g.,][]{Kaspi_2010,GW170817}. NSs are ubiquitous in the Universe, and have close connections to a number of current puzzles in astronomy, including the origins of the mysterious Fast Radio Bursts \citep[e.g.,][and references therein]{Petroff+2022}.

NSs in the form of millisecond pulsars (MSPs) and X-ray binaries are especially abundant in globular clusters\footnote{\url{http://www.naic.edu/~pfreire/GCpsr.html}}, where the MSP specific abundance is about an order of magnitude larger than the one in the Galactic field \citep{Clark_1975,Katz_1975,Ransom_2008}. It is now well understood that the frequent gravitational encounters dictated by the high stellar densities of globular clusters efficiently catalyze the dynamical formation of MSPs \citep[e.g.,][]{Bhattacharya_vandenHeuvel_1991,Sigurdsson_Phinney_1995,Ivanova+2008,Ye_msp_2019}. These past few years have seen a rapid increase in the number of detected cluster MSPs thanks to the advent of new radio telescopes, such as \textit{FAST} and \textit{MeerKAT} \citep[e.g.,][]{Pan+2021,Ridolfi+2021}. However, the true underlying pulsar population is still rather uncertain owing to selection biases, including the luminosity thresholds of surveys, the dispersion by the interstellar medium, the spreading of pulsar signals by Doppler shifting in binaries, and the beaming fraction \citep[e.g.,][for a review]{Lorimer_2008}.

Many previous studies have estimated the pulsar population in star clusters using different methods, including pulsar luminosity functions \citep{Fruchter_Goss_1990,Kulkarni+1990,Wijers+1991,Hui+2010,Bagchi+2011}, binary population synthesis with a cluster background \citep{Ivanova+2008}, combining stellar encounter rates with observations (X-ray sources, \citealp{Heinke+2005}; gamma-ray luminosities, \citealp{Abdo+2010}), and Bayesian analysis \citep[e.g.,][]{Turk_Lorimer_2013}. However, different methods have yielded very different results and no studies have explored systematically the population of cluster pulsars with self-consistent globular cluster simulations.

In the past decade, MSPs from globular clusters have been connected to the Galactic-Center gamma-ray excess detected by the \textit{Fermi}-Large Area Telescope \citep{Goodenough_Hooper_2009}. This gamma-ray excess is roughly spherical symmetric about the Galactic Center, and extends out to $\sim 2~$kpc, which cannot be explained by the cosmic ray interaction with interstellar medium or other known gamma-ray sources \citep[e.g.,][]{Murgia_2020}. The two main competing explanations for the excess are dark matter annihilation and/or an unresolved MSP population \citep[e.g.,][and references therein]{Hooper_Goodenough_2011,Hooper_Linden_2011,Abazajian_2011,Abazajian+2012,Murgia_2020}. While the former scenario is challenged by the non-detection of gamma-rays from dwarf spheroidal galaxies \citep{Murgia_2020}, the small number of low-mass X-ray binaries, which are believed to be the progenitors of MSPs \citep[e.g.,][]{Alpar+1982,Bhattacharya_vandenHeuvel_1991}, detected in the Galactic Center \citep{Cholis+2015,Haggard+2017} argues against the in-situ MSP hypothesis (although see \citealp{Gautam+2022}). Instead, previous studies have suggested that MSPs delivered by globular clusters that inspiraled into the Galactic central region may explain the excess \citep{Brandt_Kocsis2015,Abbate+2018,Fragione+2018gce}.

In this study, we systematically explore a large number of cluster models simulated with the $N$-body Monte Carlo code \texttt{Cluster Monte Carlo} (\cmc; \citealp{Kremer+2020catalog}) to calculate the number of MSPs in globular clusters with a wide range in initial conditions, including different initial masses, metallicities, and virial radii. We derive scaling relations as a function of the cluster mass and age, which we then apply to estimate the population of MSPs delivered to the Galactic Center by inspiraling globular clusters. Compared to previous studies that estimated the number of MSPs \citep[e.g.,][]{Fruchter_Goss_1990,Kulkarni+1990,Wijers+1991,Heinke+2005,Ivanova+2008,Abdo+2010,Hui+2010,Bagchi+2011,Turk_Lorimer_2013} or their gamma-ray luminosities in dense star clusters \citep[e.g.,][]{Brandt_Kocsis2015,Abbate+2018,Fragione+2018gce,Naiman+2020}, our approach is more realistic since the catalog models include a self-consistent detailed treatment of single and binary pulsar evolution in a dynamical environment, and are representative of the Milky Way globular clusters.

Our paper is organized as follows. In Section~\ref{sec:msps}, we discuss how the number of MSPs in globular clusters depends on the clusters’ properties, and provide scaling relations to estimate their number as a function of cluster mass and age. In Section~\ref{sec:gce}, we estimate the population of MSPs delivered to the Galactic Center from inspiralling globular clusters and compare their gamma-ray luminosity to the observed Galactic-Center gamma-ray excess. We discuss the implications of our finding and draw our conclusions in Section~\ref{sec:discuss_conclude}.

\section{Millisecond Pulsar Population in Globular Clusters} \label{sec:msps}

\subsection{$N$-body Models}\label{subsec:models}
We use the \cmc~cluster catalog models \citep{Kremer+2020catalog} to estimate the population of MSPs in globular clusters. These catalog models were run with the \cmc~$N$-body dynamics code \citep[e.g.,][and references therein]{Rodriguez+2021CMC}, which is based on the H\'{e}non-style orbit-averaged Monte Carlo method \citep{henon1971monte,henon1971montecluster}. The models span a wide range of initial conditions, including different initial numbers of stars ($N=2\times10^5$, $4\times10^5$, $8\times10^5$, $1.6\times10^6$), Galactocentric distances ($R_g/\rm{kpc}=2, 8, 20$), virial radii ($R_v/\rm{pc} = 0.5, 1, 2, 4$), and metallicities ($Z = 0.0002, 0.002, 0.02$). The initial stellar density distributions in the models follow a King profile \citep{King1966} with a concentration parameter $W_0=5$. The initial masses of the stars are drawn from a Kroupa initial mass function (IMF; \citealp{Kroupa2001}) between $0.08$ and $150~\msun$. All models have a $5\%$ initial binary fraction where the masses of the companion stars are drawn from a flat distribution in mass ratio to the primary star in the range $[0.1-1]$ \citep[e.g.,][]{Duquennoy_Mayor_1991}. The initial binary separations are sampled from a log-uniform distribution from near contact ($a\geq5(R_1+R_2)$, where $R_1$ and $R_2$ are the stellar radii) to the hard/soft boundary, and the initial binary eccentricities follow a thermal distribution \citep[e.g.,][]{Heggie1975}. Finally, the natal kicks of NSs formed in core-collapse supernovae are sampled from a Maxwellian distribution with a standard deviation $\sigma_{ccsn}=265\,\rm{km \,s^{-1}}$ \citep{Hobbs+2005}, while NSs formed in electron-capture supernovae and accretion-induced collapses receive smaller natal kicks drawn from a Maxwellian distribution with a standard deviation $\sigma_{ecsn} = 20\,\rm{km\,s^{-1}}$ \citep{Kiel+2008,Ye_msp_2019}. All models are evolved up to a Hubble time, and they reproduce well the observed properties of the population of Galactic globular clusters \citep[][and their Figure 2]{Kremer+2020catalog}.

In our cluster models, MSPs are modeled following \citet[][and references therein]{Ye_msp_2019}. In short, NSs are all assumed to be born as young pulsars with large magnetic fields between $10^{11.5}-10^{13.8}~$G and spin periods between $30$~ms and $1000$~ms. MSPs are formed during periods of stable mass transfer from their companion stars in binaries. As a result of Roche lobe overflow, the young pulsar or the old non-radiating NS will be spun up by the mass and angular momentum transferred from the inner edge of a Keplerian accretion disc \citep[][Eq. 54]{hurley2002evolution}, and its magnetic field will decay following
\begin{equation}
    B = \frac{B_0}{1+\frac{\Delta M}{10^{-6}\,\msun}}\exp\left(-\frac{T-t_{\rm acc}}{\tau}\right)+5\times10^7 \,\rm{G}\,,
\end{equation}
where $B_0$ is the magnetic field at the beginning of the mass transfer period, $t_{\rm acc}$ is the time spent on accretion, and $\tau=3~$Gyr is the decay timescale of the magnetic field of isolated pulsars. We assume a lower limit of $5\times10^7~$G for the MSP magnetic fields. Note that our simple prescriptions allow us to reproduce closely the observed magnetic fields and spin periods of cluster pulsars, as shown in \citet{Ye_msp_2019}.

\subsection{Gamma-ray luminosity and Millisecond Pulsar Population}\label{subsec:gammaray}
\begin{figure}
    \includegraphics[width=\columnwidth]{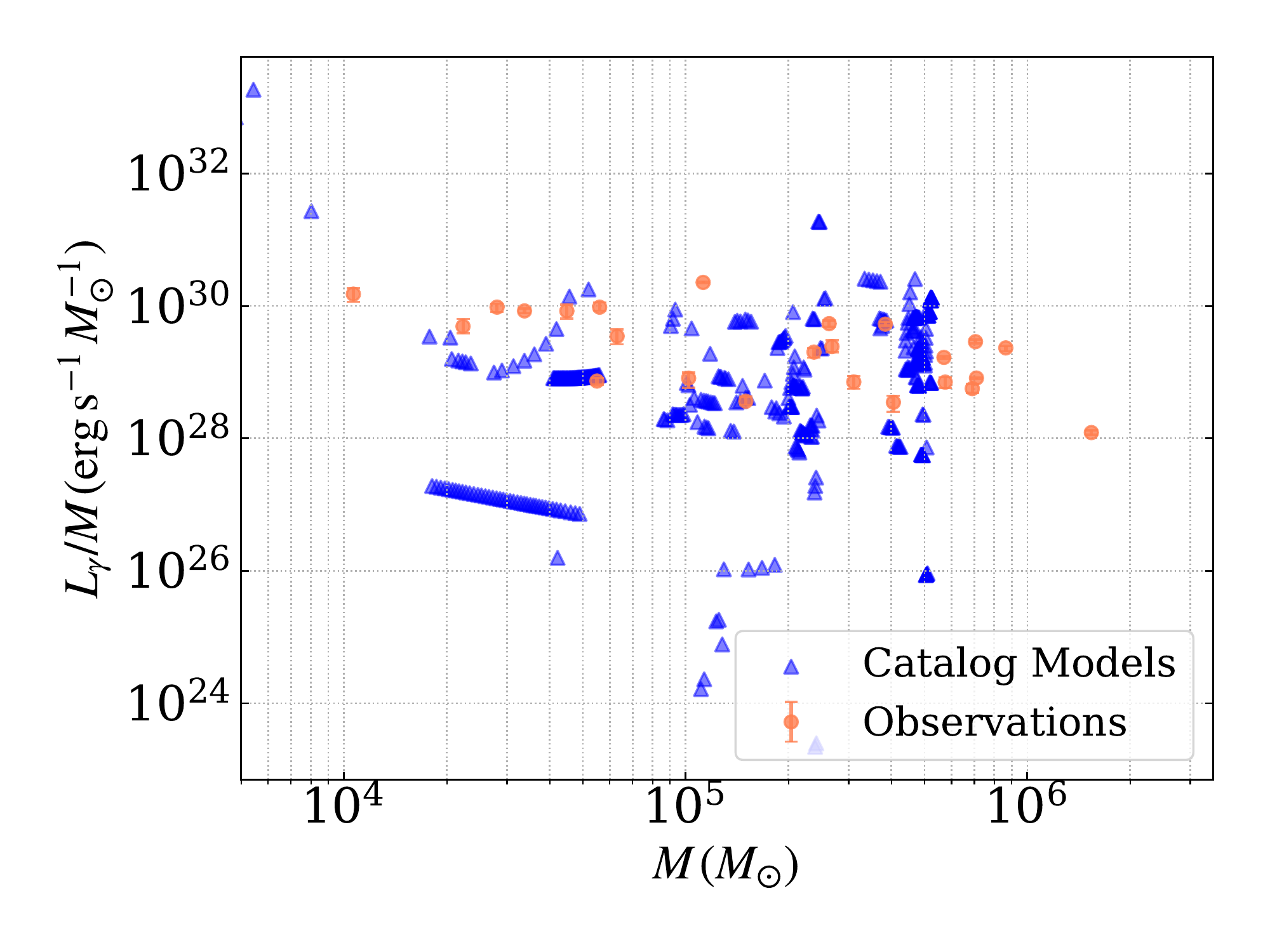}
    \caption{Gamma-ray luminosity per cluster mass as a function of the cluster mass. We show gamma-ray luminosities from multiple time steps, within 2~Gyrs until either the time of cluster disruption or a Hubble time. We only show model clusters that survived to the present day or dissolved after $7~$Gyr, which is about the age of the known youngest globular clusters in the Milky Way \citep[][and references therein]{Forbes_Bridges_2010,Dotter+2010,Dotter+2011,VandenBerg+2013,Kruijssen+2019}}
    \label{fig:Lgamma_M}
\end{figure}

The \textit{Fermi} Gamma-ray Space Telescope has detected gamma-ray emission from a number of globular clusters, generally thought to originate from MSPs therein present. To compare our models against observations, we estimate the total gamma-ray luminosity of our cluster models from their population of MSPs. The gamma-ray luminosities of model MSPs are calculated using \citep[see][Eq.~3.3]{Hooper_Mohlabeng2016}
\begin{equation}
    L_{\gamma MSP} \approx 4.8\times10^{33} \left(\frac{B}{10^{8.5}\,\rm{G}}\right)^2\left(\frac{P}{3\,\rm{ms}}\right)^{-4}\left(\frac{\eta}{0.1}\right)\,\rm{erg/s}\,,
    \label{equ:lgamma}
\end{equation}
where $B$ and $P$ are the magnetic field and spin period of a MSP, respectively, and $\eta$, which we fix to $0.1$, is the gamma-ray luminosity emission efficiency. Figure~\ref{fig:Lgamma_M} shows the observed gamma-ray luminosities per unit mass from $25$ globular clusters detected by \textit{Fermi} \citep[][Table~1]{Hooper_Linden_2016}, and model gamma-ray luminosities from $70$ catalog models with a non-zero number of MSPs (within the last 2~Gyr). We find that most of our models well overlap with the region defined by the observed population\footnote{The outliers where $L_{\gamma}/M\lesssim10^{28}\,\rm{erg\,s^{-1}}\,\msun^{-1}$ are from $\sim 10~$ models with different initial conditions. All of them except one has only one MSP (one model has four MSPs) within the last 2~Gyrs. Most of these MSPs have low gamma-ray luminosities because they were formed early in the evolution of their host clusters and evolved in isolation where their magnetic fields decayed to $\lesssim10^8~$G and they spun down slightly, or they went through more mass transfer periods and their magnetic fields were further reduced. Therefore, the offsets from the observations are from a small number of MSPs combined with their low gamma-ray luminosities.}. As already discussed, our models are consistent with the observed Galactic globular cluster properties such as their masses and half-light radii. Figure~\ref{fig:Lgamma_M} illustrates that our catalog models can also reproduce the observed gamma-ray luminosities of Galactic globular clusters quite well.

\begin{figure*}[t]
    \includegraphics[width=\textwidth]{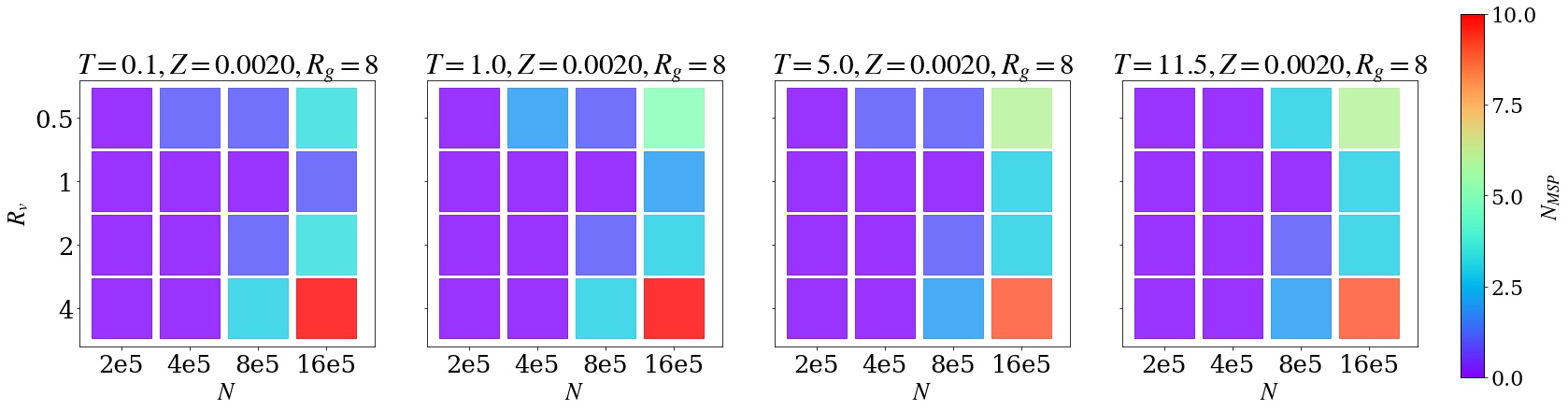}
    \caption{Numbers of MSPs for models with different initial numbers of stars $N$ and virial radii $R_v$. All models have metallicity $Z=0.002$ and Galactocentric distance $R_g = 8~$kpc. From left to right the panels show the numbers of MSPs at 0.1, 1., 5., and 11.5~Gyr, respectively.}
    \label{fig:z0.002_rg8}
\end{figure*}

To further break down the MSP number dependence on the properties of a globular cluster, we show in Figure~\ref{fig:z0.002_rg8} the numbers of MSPs for models with metallicity $Z=0.002$, Galactocentric distance $R_g=8~$kpc, and various initial numbers of stars $N$ and viral radii $R_v$. Models with different metallicities and Galactocentric distances are shown in Figure~\ref{fig:z0.0002_rg2}-\ref{fig:z0.02_rg20} in the Appendix. We find that the most massive and densest globular clusters with initial $N\ge8\times10^5$ and $R_v=0.5~$pc contribute the largest number of MSPs (see Figures~\ref{fig:z0.002_rg8} and \ref{fig:z0.0002_rg2}-\ref{fig:z0.02_rg20}). This is expected since the more massive and denser a globular cluster is, the higher is its dynamical interaction rate, which is the key process to form MSPs \citep{Bhattacharya_vandenHeuvel_1991,Sigurdsson_Phinney_1995,Ivanova+2008,Ye_msp_2019}.

\subsection{Scaling Relations}\label{subsec:fittings}
\begin{figure}[h]
    \includegraphics[width=\columnwidth]{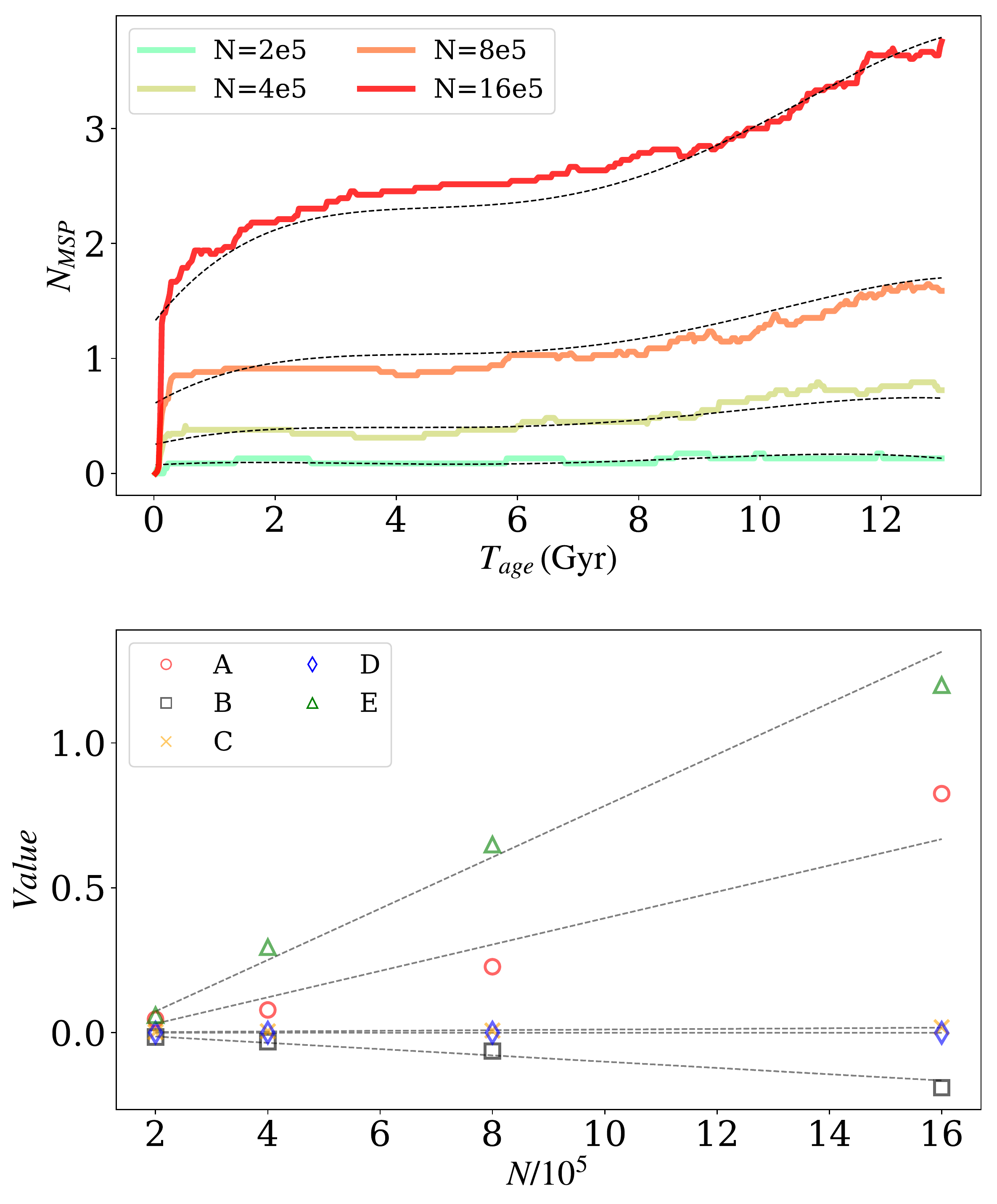}
    \caption{Top panel: The average number of MSPs for models with different initial number of stars, $N$, as a function of the cluster age. The solid curves represent the model averages, while the black dashed curves show the fits from Eq.~\ref{equ:one}. Bottom panel: the markers and the black dashed lines show the polynomial parameters and their fits from Eq.~\ref{equ:one}, respectively.}
    \label{fig:all_fit}
\end{figure}

We calculate the average number of MSPs for catalog models that survived to the present day \citep[see Table~6 in][for surviving globular clusters]{Kremer+2020catalog}, and express the averages as a function of the cluster age with a polynomial function. The polynomial parameters can be taken to be linear functions of the initial numbers of cluster stars or the initial cluster masses $M$, where $N=M/0.6\msun$ for a canonical Kroupa IMF. Our polynomial fits comparing to the model data are shown in Figure~\ref{fig:all_fit}. The polynomial fits as a function of the cluster age and the initial cluster mass are as follows

\begin{equation}\label{equ:one}
    \begin{aligned}
        N_{MSP} &= A \times t + B \times t^2 +C \times t^3 + D \times t^4 + E,\\
        A &= 0.046 \times M_5 - 0.060,\\
        B &= -0.011 \times M_5+0.008,\\
        C &= 0.0011 \times M_5-0.0003,\\
        D &= -0.00003 \times M_5-0.00001,\\
        E &= 0.089 \times M_5-0.104,
    \end{aligned}
\end{equation}
where $M_5 = M/(0.6\times10^5 \msun)$ and $t$ is in unit of Gyr. Note that the number of MSPs for different $N$ (or $M$) could have large fluctuations (see Figures~\ref{fig:z0.002_rg8} and  \ref{fig:z0.0002_rg2}-\ref{fig:z0.02_rg20}) because of the different initial conditions and small number statistics. The standard deviations can be up to a factor of about $2$ the average values for dense star cluster models with $N>2\times10^5$.

The initial cluster mass can be estimated from the present-day mass taking into account the fact that most catalog models that survived to the present day lost $\sim 50-60\%$ of their initial mass depending on their galactocentric distance (in our models the mean value is about $60\%$ and the median is about $55\%$; the mass loss rate for models with $R_g=2~$kpc is about $10\%$ larger for both values)\footnote{The catalog models assume a point-mass spherical Galactic potential, and do not take into account the tidal shocking effect of the Galactic disk, so they potentially underestimate the mass loss rate from this effect.}. As an example, we estimate the number of MSPs in the globular cluster 47 Tucanae, assuming its present-day mass to be $\sim 10^6\msun$ (\citealp[][2010 edition]{Harris_1996}; \citealp{Baumgardt_Hilker2018,Ye_47tuc_2021}), and that it lost about half of its mass in $\sim 11$~Gyr. For an initial cluster mass of $2\times10^6\,\msun$, the number of MSPs estimated from Eq.~\ref{equ:one} at 11~Gyr is about 9. We use a factor of about $2$ for the $1\sigma$ upper limit and take into account that $\sim 70\%$ of its MSPs may come from binary formation through giant star collisions with NSs and tidal capture interactions between a NS and a main-sequence star \citep{Ye_47tuc_2021}. The final estimated number of MSPs in 47 Tucanae at the present day is $\sim 60$, consistent with the estimates from the previous targeted simulation of the cluster \citep{Ye_47tuc_2021} and with the observations \citep[e.g.,][]{Heinke+2005,Abdo+2009}. This example also demonstrates that our scaling relations can be used to estimate reasonably the number of MSPs in clusters with masses larger than the ones in the \cmc~catalog models.

\section{Application to the Galactic-Center Gamma-Ray Excess}\label{sec:gce}
In this Section, we adopt the semi-analytical method described in \citet{Gnedin+2014} for building a Galactic potential, sampling an initial population of globular clusters, and inspiraling the clusters through dynamical friction into the Galactic Center (Section~\ref{subsec:method}). We then use the masses of the inspiraled globular clusters to estimate the number of MSPs delivered to the Galactic Center and their gamma-ray emission (Section~\ref{subsec:msp_gc}). This semi-analytical approach allows us to sample a large number of globular clusters without significant computational costs.

For the purpose of estimating the number of MSPs in the Galactic Center from inspiraling globular cluster, we also calculate the average MSPs for all models with initial $R_g=2~$kpc (including dissolved models). Here, we only consider models with $R_g=2~$kpc because they can most closely represent the inspiraling globular clusters affected strongly by the Galactic potential.  The polynomial fits as a function of the cluster age and the initial cluster mass are as follows

\begin{equation}\label{equ:two}
    \begin{aligned}
    N_{MSP} &= A_{rg2} \times t + B_{rg2} \times t^2 +C_{rg2} \times t^3 + D_{rg2},\\
    A_{rg2} &= 0.018 \times M_5 - 0.009,\\
    B_{rg2} &= -0.0030 \times M_5+0.0015,\\
    C_{rg2} &= 0.0002 \times M_5-0.0002,\\
    D_{rg2} &= 0.068 \times M_5-0.157.
    \end{aligned}
\end{equation}

\begin{figure}
    \includegraphics[width=\columnwidth]{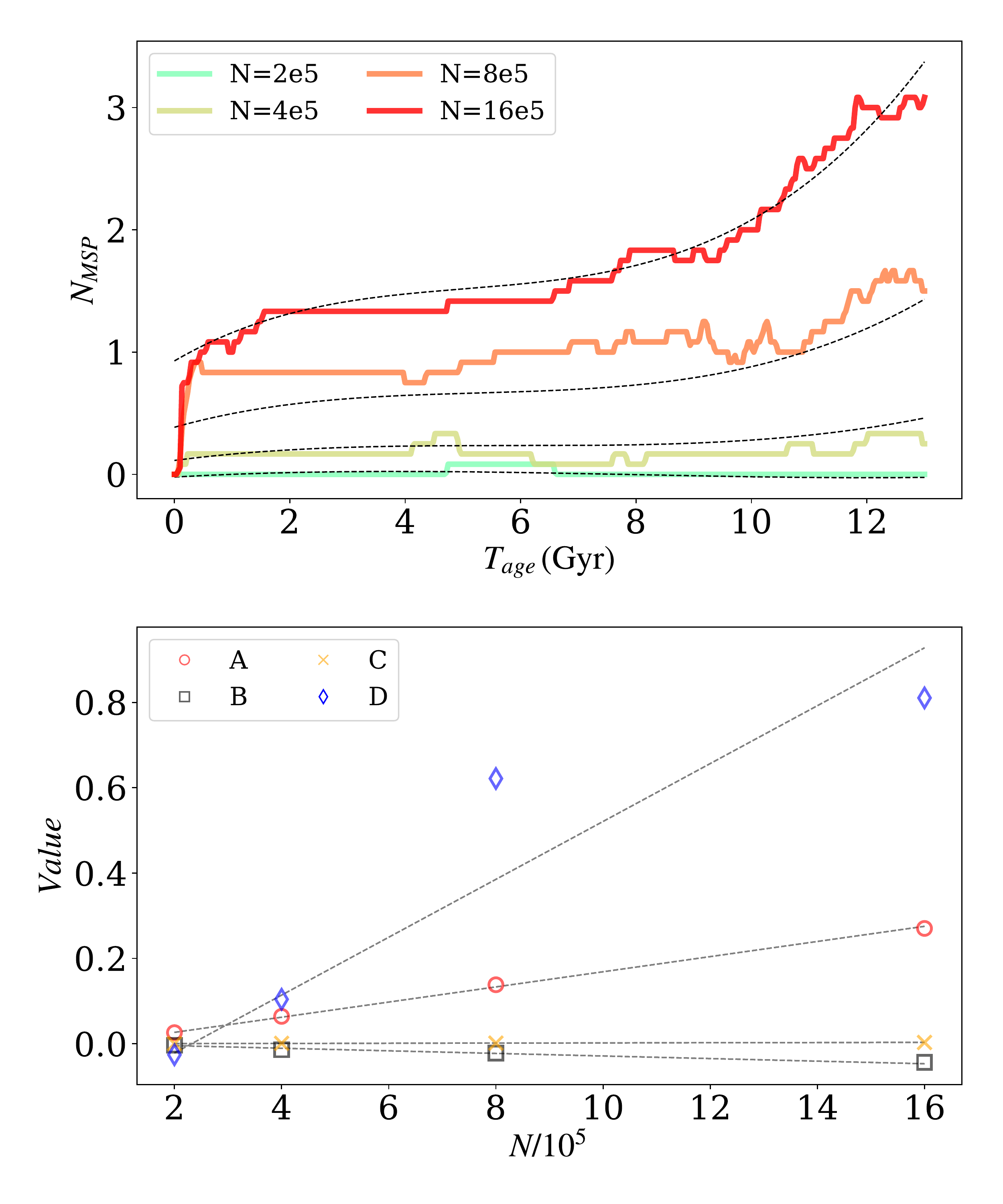}
    \caption{Similar to Figure~\ref{fig:all_fit}, but only for models with a Galactocentric distance $R_g=2~$kpc.}
    \label{fig:rg2_fit}
\end{figure}

Figure~\ref{fig:rg2_fit} compares the fits to the model data. When compared to the fits using all models as in Figure~\ref{fig:all_fit}, we find that there could be larger deviations, especially for initial $N=8\times10^5$ in Figure~\ref{fig:rg2_fit} (top panel). This is probably due to the fact that the number of models is smaller (48 models with $R_g=2~$kpc compared to 119 total non-dissolved models). Note that we include both not-yet-dissolved and non-dissolved models in this calculation.

\subsection{Semi-analytical Methods} \label{subsec:method}
We briefly summarize the semi-analytical methods from \citet{Gnedin+2014}. We assume that the Galaxy is composed of stars whose distribution follows a spherical S\'ersic mass density profile \citep[e.g.,][]{Terzic+2005}, and a dark matter halo with an Navarro–Frenk–White profile \citep{BT_galacticdynamics}. The S\'ersic profile has a total mass $M_s=5\times10^{10}\,\msun$, a concentration index $n_s=2.2$, and an effective radius $R_s=4$~kpc. The Navarro–Frenk–White profile has a total mass $M_h=10^{12}\,\msun$, and a scale radius $R_h=20$~kpc. In addition, we also include a central supermassive black hole with a mass $M_{SMBH} = 4\times10^6\,\msun$.

To initialize a population of globular clusters, we assume that they follow the mass distributions of the stars in the Galaxy, and their total mass is a fixed fraction, $1.2\%$, of the field stars \citep{Gnedin+2014}. We also assume that clusters are on circular orbits, and sample their initial semi-major axes in the Galaxy between $0.1$ and $100$ kpc. The individual masses of the globular clusters are drawn from a power-law distribution
\begin{equation}
    \frac{dN_{GC}}{dM_{GC}} \propto M_{GC}^{-2},
\end{equation}
where $N_{GC}$ is the number of globular clusters at a certain mass, and $M_{GC}$ is the cluster mass, in the range $10^4\,\msun$-$10^7\,\msun$. Finally, we adopt the average density at the half-mass radius
\begin{equation}
\rho_{\mathrm{h}}=10^3\min\left\{10^2,\max\left[1,\left(\frac{M}{10^5\ \msun}\right)^2\right]\right\}\,\frac{\msun}{\mathrm{pc}^3}.
\label{eqn:rhalfm}
\end{equation}

Dynamical friction leads to the gradual inspiral of the globular clusters in the Galactic potential. We calculate the evolution of a cluster's distance to the Galactic center following \citep{BT_galacticdynamics,Gnedin+2014}
\begin{equation}
    \frac{dr^2}{dt} = -\frac{r^2}{t_{df}},
\end{equation}
\begin{equation}
    t_{df} \approx 0.23 \left(\frac{r}{\rm{kpc}}\right)^2 \left(\frac{M_{GC}}{10^5 \msun}\right)^{-1} \left(\frac{v_c}{\rm{km\,s^{-1}}}\right)\,\rm{Gyr}\,,
\end{equation}
where $v_c$ is the circular orbital velocity of a globular cluster.

During inspiral, a globular cluster will lose mass from the stripping of stars by the Galactic tidal field, evaporation of stars through two-body relaxation, and stellar evolution. The timescale of mass loss from the stripping of stars by the Galactic tidal field can be estimated by  \citep{Gieles_Baumgardt_2008,Gnedin+2014}
\begin{equation}
    t_{tid} \approx 10\left(\frac{M_{GC}}{2\times10^5\msun}\right)^{2/3}P(r)\, \rm{Gyr}\,,
\end{equation}
where
\begin{equation}
    P(r) = 41.4\left(\frac{r}{\rm{kpc}}\right)\left(\frac{v_c}{\rm{km\,s^{-1}}}\right)^{-1}\,.
\end{equation}
For isolated globular clusters, the evaporation time follows \citep[][and references therein]{Gnedin+2014}
\begin{equation}
    t_{iso} = 17\left(\frac{M_{GC}}{2\times10^5\msun}\right)\,\rm{Gyr}\,.
\end{equation}
The mass loss rate of a globular cluster from these two effects is
\begin{equation}
    \frac{dM_{GC}}{dt} = -\frac{M_{GC}}{\min(t_{iso}, t_{tid})}.
\end{equation}
Typically, $t_{tid}<t_{iso}$ in the inner regions of the galaxy, which harbor the globular clusters that can potentially spiral into the Galactic Center. Finally, to incorporate mass loss from stellar evolution, we assume a broken-power-law IMF as in \citet{Kroupa2001}, with stellar masses between $0.08$ and $150\,\msun$. We model the initial-to-final mass relation following \citet[][Eq.~7.22]{Merritt_2013}, and the turn-off mass at a given time is approximated using $m_{\rm TO}\approx(t_{ms}/ 10\,{\rm Gyr})^{-2/5}$ \citep{Hansen_Kawaler_1994}. Note that the mass lost from the globular clusters is added to the field stellar mass at each time step.

We evolve our population of globular clusters for $11.5$ Gyr assuming that all clusters formed from a burst of star formation at redshift $z=3$ \citep{Fragione+2018gce}. We classify globular clusters as disrupted when the cluster mass densities at their half-mass radii are smaller than the surrounding Galactic field density, when $M_{GC} < 100 \msun$ (\citealp[][2010 edition]{Harris_1996}; \citealp{Baumgardt_Hilker2018}, also see Figure \ref{fig:ngc_mgc} below), or when its distance with respect to the Galactic Center is smaller than $1$~pc.

\subsection{Millisecond Pulsars and the Gamma-ray Excess in the Galactic Center}\label{subsec:msp_gc}
\begin{figure}
    \includegraphics[width=\columnwidth]{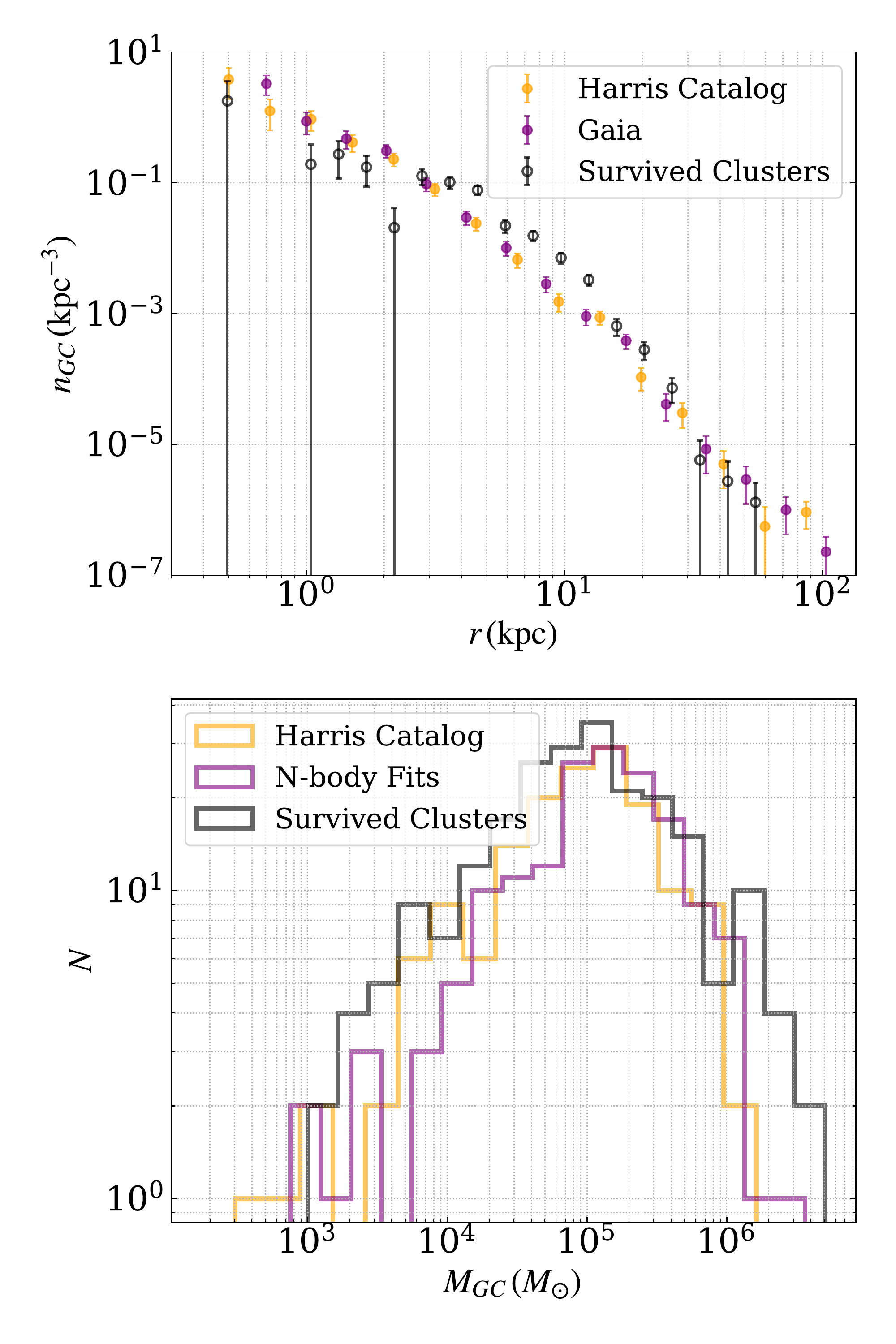}
    \caption{Top panel: Number density of globular clusters as a function of the Galactocentric distance in the Milky Way. The orange and purple dots show the observations from the Harris catalog \citep[][2010 edition]{Harris_1996} and the Gaia EDR3 \citep{Baumgardt+2021}, repsectively. The black dots show the survived globular clusters at 11.5 Gyr from our cluster samples. Bottom panel: Mass distribution of globular clusters in the Milky Way. The orange histogram is from the Harris catalog \citep[][2010 edition]{Harris_1996} assuming a mass-to-light ratio of 1.5. The purple histogram is from $N$-body fits to the observed cluster surface brightness profiles and velocity dispersion profiles \citep{Baumgardt_2017,Baumgardt_Hilker2018}. The mass distribution of the survived sample clusters are shown in the black histogram.}
    \label{fig:ngc_mgc}
\end{figure}

We sample about $8700$ globular clusters with an initial total mass of about $5.4\times10^8\,\msun$. We find that after $11.5$~Gyr most of the globular clusters are disrupted, with only about $200$ systems surviving to the present day, which is in nice agreement with the number of observed globular clusters in our Galaxy \citep[][2010 edition]{Harris_1996}. Figure~\ref{fig:ngc_mgc} compares the number density distribution and mass distribution of the survived sample globular clusters to the observations. The results from our simple semi-analytical method are consistent with most of the features in the observed population of Galactic globular clusters. In addition, since our semi-analytical method is based on the method from \citet[][and references therein]{Gnedin+2014}, where it was shown that the method reproduces broadly the properties (including masses and radii) of Milky Way globular clusters, the survived sample clusters should also have half-mass radii consistent with the observations.

We use Eq.~\ref{equ:two} to estimate the number of MSPs delivered to the Galactic Center by inspiraling globular clusters\footnote{We assume that all MSPs formed in a globular cluster that are not ejected stay close to the cluster center and do not get stripped during the gradual disruption of the outermost regions of their host globular clusters.}. In addition, a non-negligible number of MSPs are ejected from the globular clusters, where $\sim 90\%$ of them are ejected at $\lesssim 100$~Myr\footnote{There are also $\sim 15$ NS-white dwarf and NS-main sequence star binaries ejected per cluster, some of which may become MSPs in the future. For simplicity we do not consider these systems.}. The number of ejected MSPs scales with the initial number of cluster stars $N$ as $N_{\rm MSP, ej}\approx \log_2(N/2\times10^5)$. Therefore, for simplicity, we add an ejected number of MSPs at the initial position of a globular cluster in the sample.

\begin{figure}
    \includegraphics[width=\columnwidth]{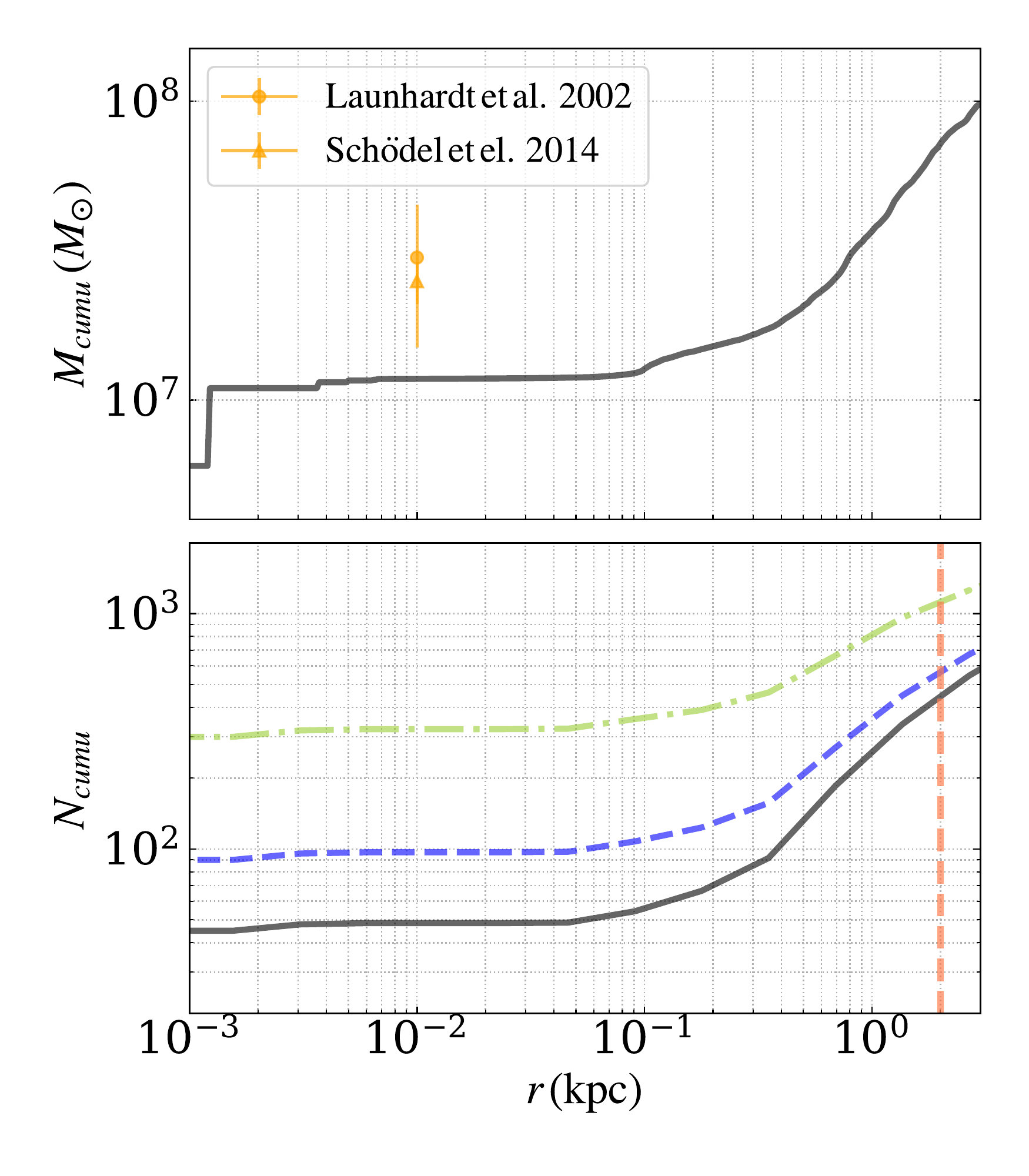}
    \caption{Top panel: Cumulative distribution of mass lost from the inspiraled globular clusters and mass from the remnants of the disrupted globular clusters. The orange markers show the observational constraints on the mass of the Milky Way nuclear star cluster \citep{Launhardt+2002,Schodel+2014a,Schodel+2014b}. Bottom panel: Cumulative number of MSPs from the disrupted globular clusters, including MSPs that were ejected initially. The black curve is calculated from the averages of Eq.~\ref{equ:two}. The blue dashed curve is the $1\sigma$ upper limit from Eq.~\ref{equ:one}, while the green dot-dashed curve shows an extreme upper limit taking into account that $70\%$ of the MSPs in a cluster can be formed from giant star collisions with NSs and tidal capture interactions (in addition to the $1\sigma$ upper limit), which were not considered in the catalog models. The vertical line marks $2$~kpc.}
    \label{fig:cumu_msp_mgc}
\end{figure}

The number of MSPs brought by inspiraled globular clusters to the Galactic Center is shown in Figure~\ref{fig:cumu_msp_mgc}. Note that the stellar mass contributed by our sample of inspiraled globular clusters are within the limits of the observed masses of the Milky Way nuclear star cluster, whose mass is equally contributed to from local star formation and inspiralled star clusters (\citealp{Launhardt+2002,Schodel+2014a,Schodel+2014b}; top panel). We find that the number of MSPs within $2$~kpc of the Galactic Center is about $400$, with a $1\sigma$ upper limit (a factor of 2 in Eq.~\ref{equ:one} as mentioned in Section~\ref{subsec:fittings}) of about $500$. While taking into account binary formation through giant star collisions with NSs and tidal capture of main-sequence stars by NSs, and assuming that $70\%$ of MSPs are formed from these channels \citep{Ye_47tuc_2021}, the number of MSPs can be up to $\sim 1000$ (also taking into account the $1\sigma$ upper limit). However, it is important to point out that this is an optimistic upper limit since the $70\%$ used is for a cluster like 47~Tucanae, which is one of the most massive and densest globular clusters in the Milky Way, and is likely to have more dynamical interactions than typical globular clusters \citep{Ye_47tuc_2021}. Note that these numbers are consistent with the numbers predicted by gamma-ray luminosity functions \citep{Dinsmore_Slatyer_2021}.

\begin{figure}
    \includegraphics[width=\columnwidth]{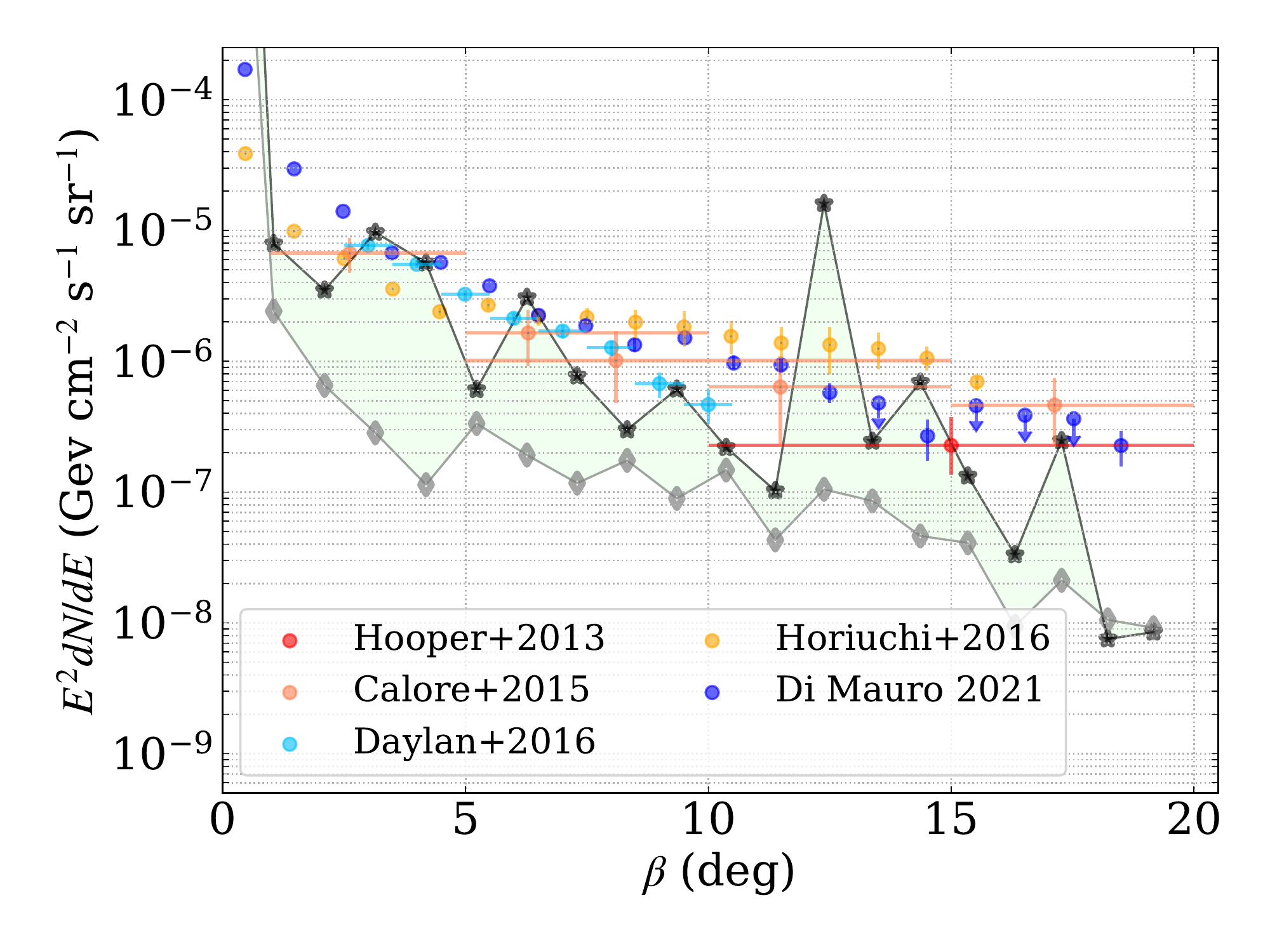}
    \caption{Gamma-ray surface brightness of MSPs from disrupted globular clusters and the gamma-ray excess detected around the Galactic Center as a function of the angular distance to the Galactic Center \citep{Hooper_Slatype_2013,Calore+2015a,Daylan+2016,Horiuchi+2016,DiMauro2021}. The gamma-ray surface brightness of MSPs corresponds to the average number of MSPs shown by the black curve in Figure~\ref{fig:cumu_msp_mgc} (bottom panel). Black stars show the case where the MSP gamma-ray luminosities are calculated using Eq.~\ref{equ:lgamma}, while gray diamonds assume that all MSPs have a gamma-ray luminosity of $4.8\times10^{33}\,\rm{erg\,s^{-1}}$.}
    \label{fig:lgev_degree}
\end{figure}

In Figure~\ref{fig:lgev_degree}, we show the gamma-ray surface brightness from the population of MSPs from inspiraled globular clusters, compared to the detected Galactic-Center gamma-ray excess \citep{Hooper_Slatype_2013,Calore+2015a,Daylan+2016,Horiuchi+2016,DiMauro2021}. The gamma-ray surface brightness corresponds to the average number of MSPs estimated from disrupted clusters (the black curve in the bottom panel of Figure~\ref{fig:cumu_msp_mgc}). The gamma-ray luminosity of each MSP is drawn randomly from a list of MSPs from all catalog models with initial $R_g = 2~$kpc, depending on when the host cluster is disrupted or when the MSP is ejected. The large fluctuation (the black star at $\gtrsim 10^{-5}\,\rm{GeV}\,cm^{-2}\,s^{-1}\,sr^{-1}$) is resulted from random draws of MSPs with large gamma-ray luminosities. The average luminosity of these MSPs is $\sim3\times10^{34}\,\rm{erg\,s^{-1}}$. For comparison, we also show in the figure the surface brightness profile when all MSPs have a gamma-ray luminosity of $4.8\times10^{33}\,\rm{erg\,s^{-1}}$ (gray diamonds; see also Eq.~\ref{equ:lgamma}). We find that, for high gamma-ray luminosities $\sim 10^{34}\,\rm{erg\,s^{-1}}$ per MSP, MSPs from inspiraled globular clusters can potentially explain most of the detected excess. However, if MSPs have lower luminosities $\sim 10^{33}\,\rm{erg\,s^{-1}}$, they may only contribute to $\lesssim 10\%$ of the detected gamma-ray excess, consistent with the previous population synthesis study \citep{Gonthier+2018}. For the extreme case where $70\%$ of cluster MSPs come from the giant star collision and tidal capture channels, and including a factor of 2 for the $1\sigma$ upper limit, MSPs from disrupted globular clusters produce a gamma-ray surface brightness comparable to the observed gamma-ray excess if all MSPs have gamma-ray luminosity $\sim 10^{34}\,\rm{erg\,s^{-1}}$.

\newpage
\section{Discussions and Conclusions}\label{sec:discuss_conclude}
To summarize, we have explored in this study how the initial conditions of dense star clusters affect the number of MSPs they produce using the \cmc\, cluster catalog models \citep{Kremer+2020catalog}. We have presented scaling relations, which express the numbers of MSPs as a function of the cluster age for different initial cluster masses in Eq.s~\ref{equ:one}-\ref{equ:two}. Our scaling relations allow quick estimates of the number of MSPs in dense star clusters.

We have also demonstrated that our model MSPs can reproduce the gamma-ray luminosities observed from globular clusters. We have applied the scaling relation we have derived to estimate the number of MSPs delivered to the Galactic Center from inspiraled globular clusters. On average, about $400$ MSPs are brought to the Galactic Center by globular clusters, with an optimistic upper limit $\sim 1000$. These MSPs have an average gamma-ray luminosity $\sim 3 \times 10^{34}\,\rm{erg\,s^{-1}}$, and can potentially explain most of the detected gamma-ray excess from the Galactic Center.

There are a few uncertainties on the predicted number of MSPs from the catalog models. We have briefly mentioned one uncertainty in Section~\ref{subsec:fittings}; that is, for massive and dense globular clusters, dynamical binary formation through NS-giant star collisions or tidal capture interactions may also contribute to a large portion of cluster MSPs \citep[$\sim 70\%$ for massive and dense globular clusters similar to 47 Tucanae, see][]{Ye_47tuc_2021}. Furthermore, observations have shown that core-collapsed globular clusters contain more isolated MSPs, which could be formed from tidal disruption events between a NS and a main sequence star \citep{Kremer_nstde_2022}. These effects combined together could contribute to a factor of a few in the number of MSPs formed in dense globular clusters, most of which are core-collapsed. However, only around $20\%$ of the present-day globular clusters are core-collapsed \citep[][2010 edition]{Harris_1996}, and if the fraction was similar for all clusters ever formed in the Milky Way, the boost from these dynamical interactions may not be very significant. In addition, we have only studied models with an initial binary fraction of $5\%$, but the rates of binary-mediated dynamical encounters could be significantly affected by different initial binary fractions. The catalog models also only adopt a standard Kroupa IMF \citep{Kroupa2001}. However, studies have suggested that the IMF of globular clusters may not be universal \citep[e.g.,][]{Marks+2012,Haghi+2017,Sollima_Baumgardt_2017,Cadelano+2020,Henault-Brunet+2020}, with different IMFs leading to different number of MSPs dynamically assembled in star clusters. For example, a top-heavy IMF (where there are more massive stars than the standard Kroupa IMF) will produce more NSs and black holes, but more black holes on average lead to fewer dynamical interactions for NSs, therefore likely fewer MSPs \citep{Ye_msp_2019}. Similarly, a bottom-light IMF will produce more black holes per cluster mass, thus may also lead to few MSPs. Finally, the predicted number of MSPs in the Galactic Center is also limited by models that are only at the constant Galactocentric distance $R_g=2~$kpc. In reality, orbits of globular clusters in a galaxy are not always circular \citep{Vasiliev+2021}, and the galactic potential they are subjected to are time-dependant during inspiral. We leave the detailed exploration of the previous uncertainties to future work \citep{Yeinprep}.

\begin{acknowledgments}
We thank Fred Rasio, Kyle Kremer, Carl Rodriguez, and the anonymous referee for useful discussions and comments. This work was supported by NSF Grants AST-1716762, AST-2108624 at Northwestern University, and by the Natural Sciences and Engineering Research Council of Canada (NSERC) DIS-2022-568580. G.F.\ acknowledges support from NASA Grant 80NSSC21K1722. This research was supported in part through the computational resources and staff contributions provided for the Quest high performance computing facility at Northwestern University, which is jointly supported by the Office of the Provost, the Office for Research, and Northwestern University Information Technology.
\end{acknowledgments}

\software{\texttt{CMC} \citep{Joshi_2000,Joshi_2001,Fregeau_2003, fregeau2007monte, Chatterjee_2010,Umbreit_2012,Chatterjee_2013b,Pattabiraman_2013,Morscher+2015,Rodriguez+2016million,Rodriguez+2021CMC}, \texttt{Fewbody} \citep{fregeau2004stellar}, \texttt{COSMIC} \citep{cosmic}, \texttt{BSE} \citep{hurley2002evolution}, \texttt{SSE} \citep{hurley2000comprehensive}}

\appendix
Similar to Figure~\ref{fig:z0.002_rg8}, we show here the number of MSPs in the catalog models with all the combinations of different initial conditions.

\begin{figure*}[h]
    \includegraphics[width=\textwidth]{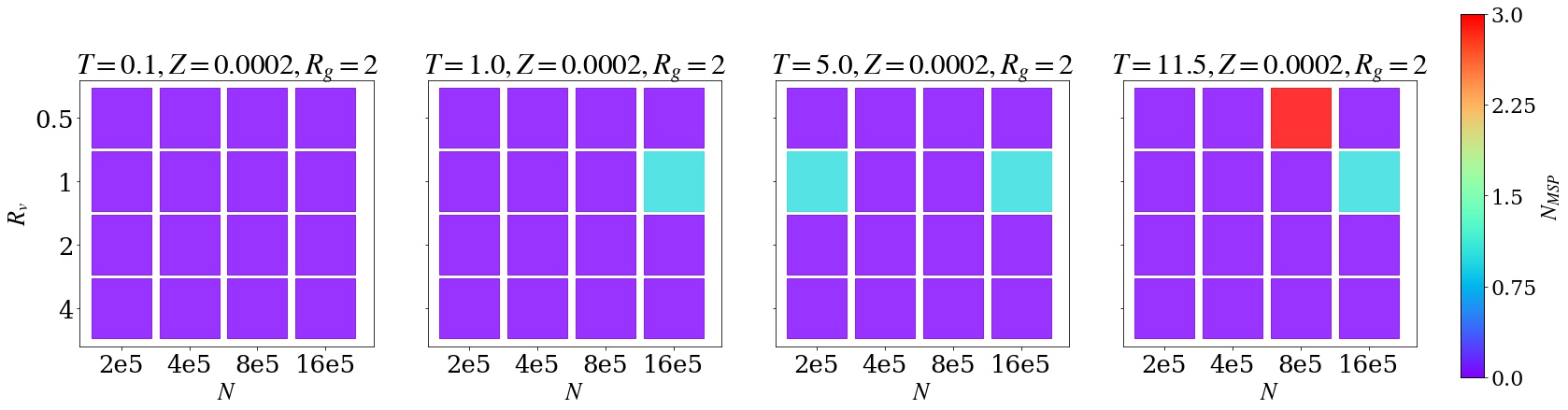}
    \caption{Same as Figure~\ref{fig:z0.002_rg8} but for models with a metallicity $Z=0.0002$ and Galactocentric distance $R_g = 2~$kpc.}
    \label{fig:z0.0002_rg2}
\end{figure*}

\begin{figure*}[h]
    \includegraphics[width=\textwidth]{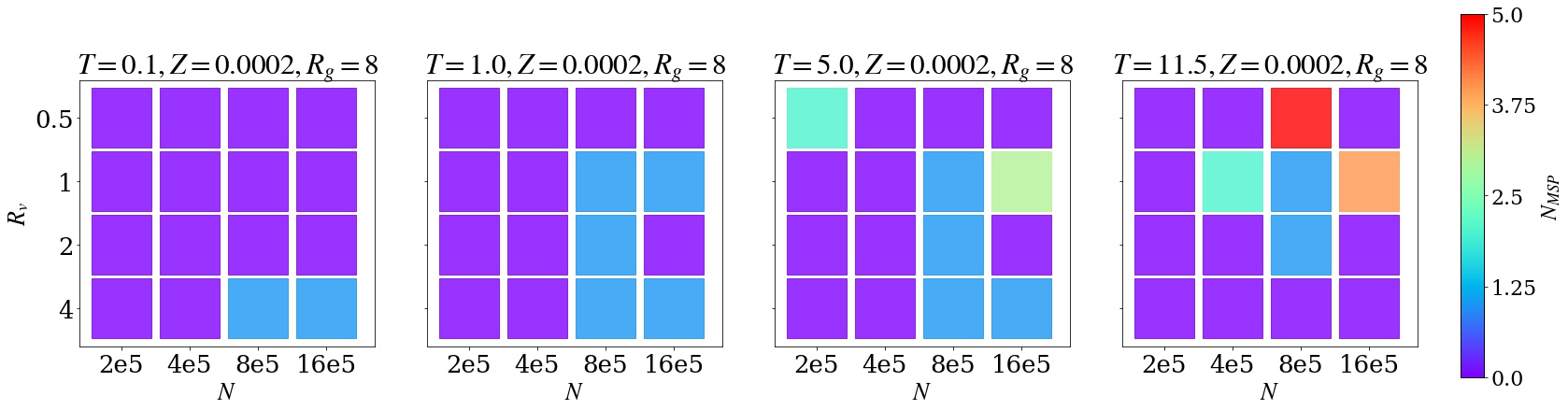}
    \caption{Same as Figure~\ref{fig:z0.002_rg8} but for models with a metallicity $Z=0.0002$ and Galactocentric distance $R_g = 8~$kpc.}
    \label{fig:z0.0002_rg8}
\end{figure*}

\begin{figure*}[h]
    \includegraphics[width=\textwidth]{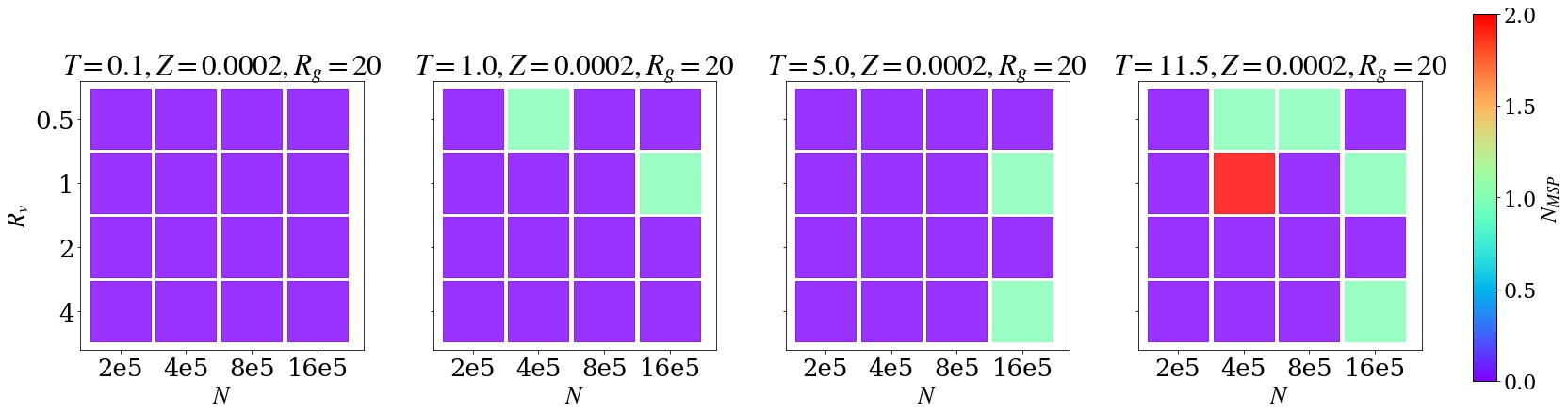}
    \caption{Same as Figure~\ref{fig:z0.002_rg8} but for models with a metallicity $Z=0.0002$ and Galactocentric distance $R_g = 20~$kpc.}
    \label{fig:z0.0002_rg20}
\end{figure*}

\begin{figure*}[h]
    \includegraphics[width=\textwidth]{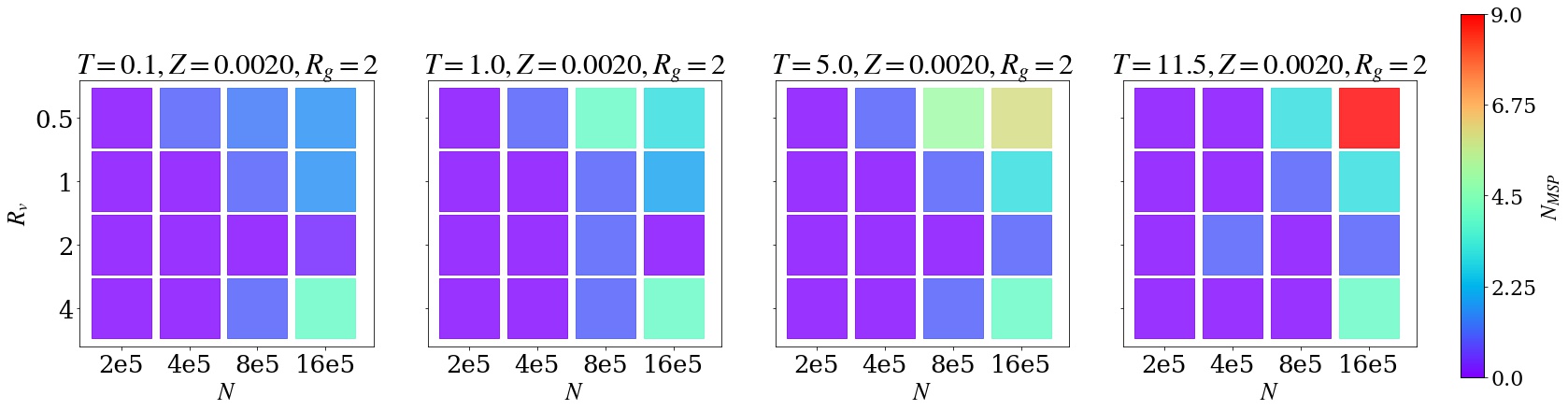}
    \caption{Same as Figure~\ref{fig:z0.002_rg8} but for models with a metallicity $Z=0.002$ and Galactocentric distance $R_g = 2~$kpc.}
    \label{fig:z0.002_rg2}
\end{figure*}

\begin{figure*}[h]
    \includegraphics[width=\textwidth]{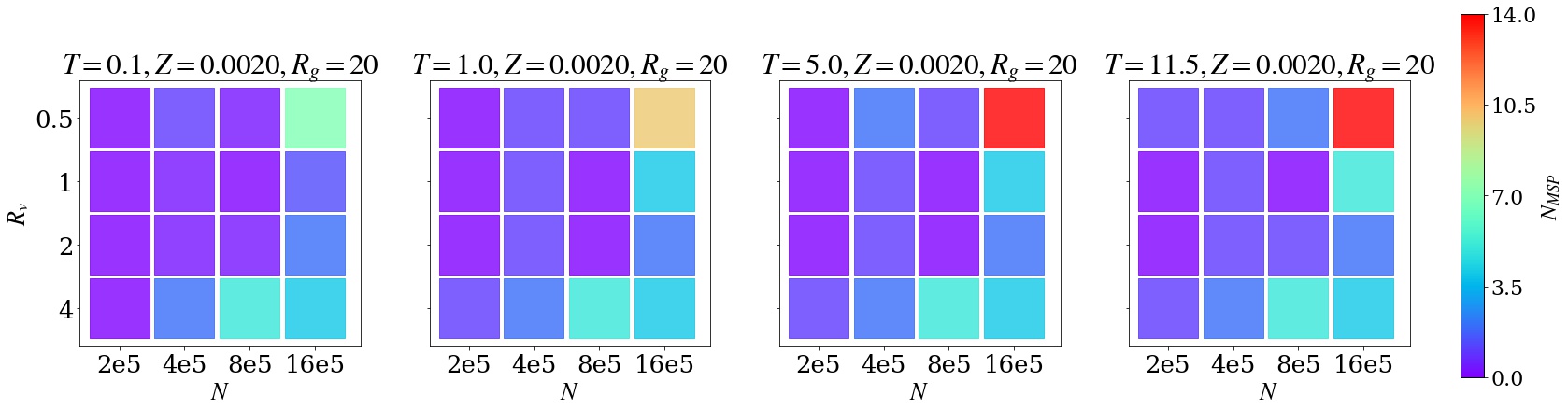}
    \caption{Same as Figure~\ref{fig:z0.002_rg8} but for models with a metallicity $Z=0.002$ and Galactocentric distance $R_g = 20~$kpc.}
    \label{fig:z0.002_rg20}
\end{figure*}

\begin{figure*}[h]
    \includegraphics[width=\textwidth]{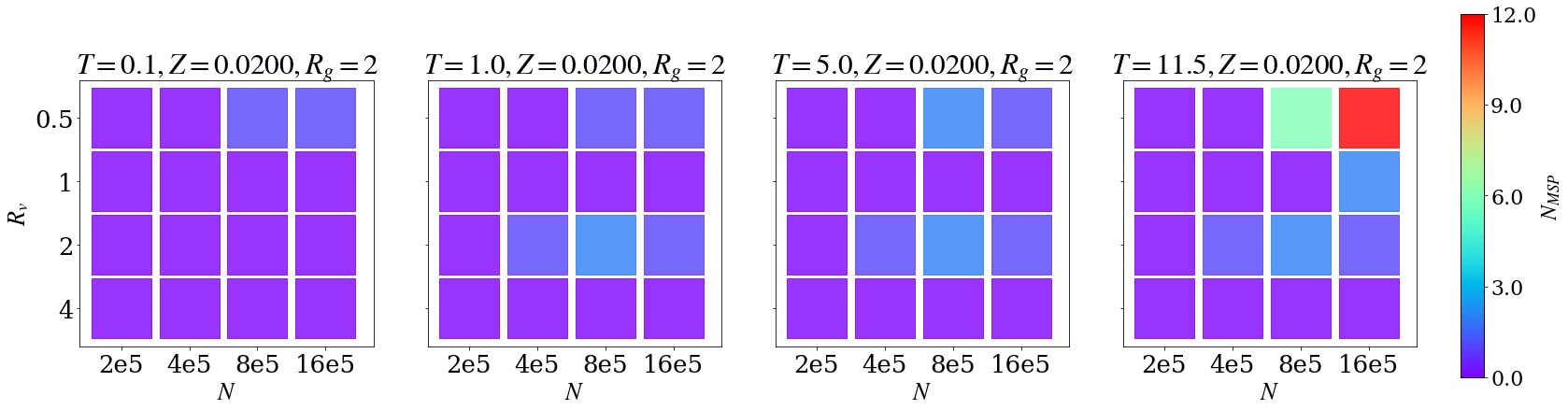}
    \caption{Same as Figure~\ref{fig:z0.002_rg8} but for models with a metallicity $Z=0.02$ and Galactocentric distance $R_g = 2~$kpc.}
    \label{fig:z0.02_rg2}
\end{figure*}

\begin{figure*}[h]
    \includegraphics[width=\textwidth]{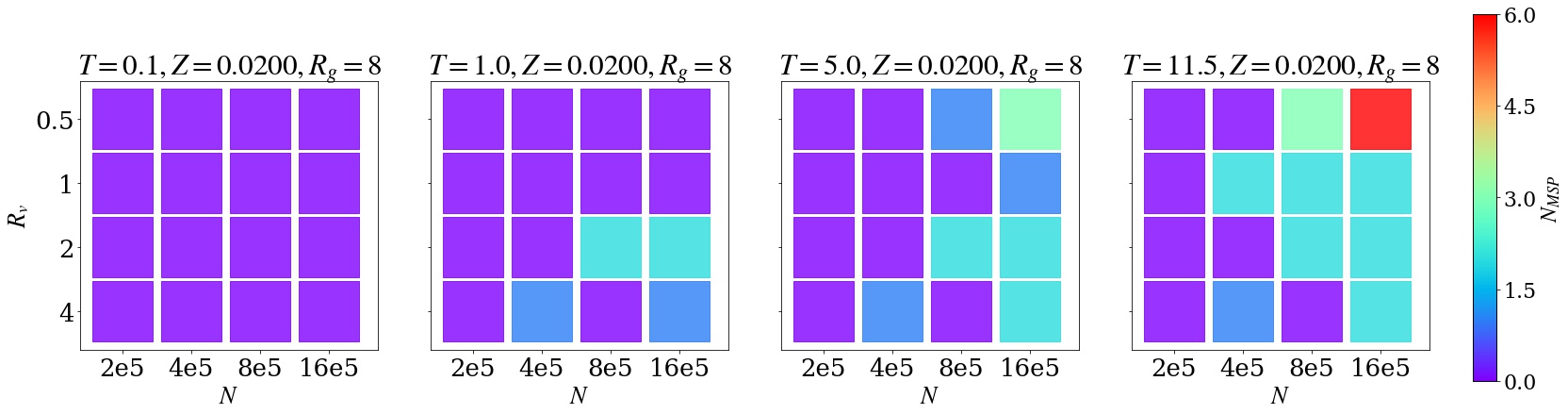}
    \caption{Same as Figure~\ref{fig:z0.002_rg8} but for models with a metallicity $Z=0.02$ and Galactocentric distance $R_g = 8~$kpc.}
    \label{fig:z0.02_rg8}
\end{figure*}

\begin{figure*}[h]
    \includegraphics[width=\textwidth]{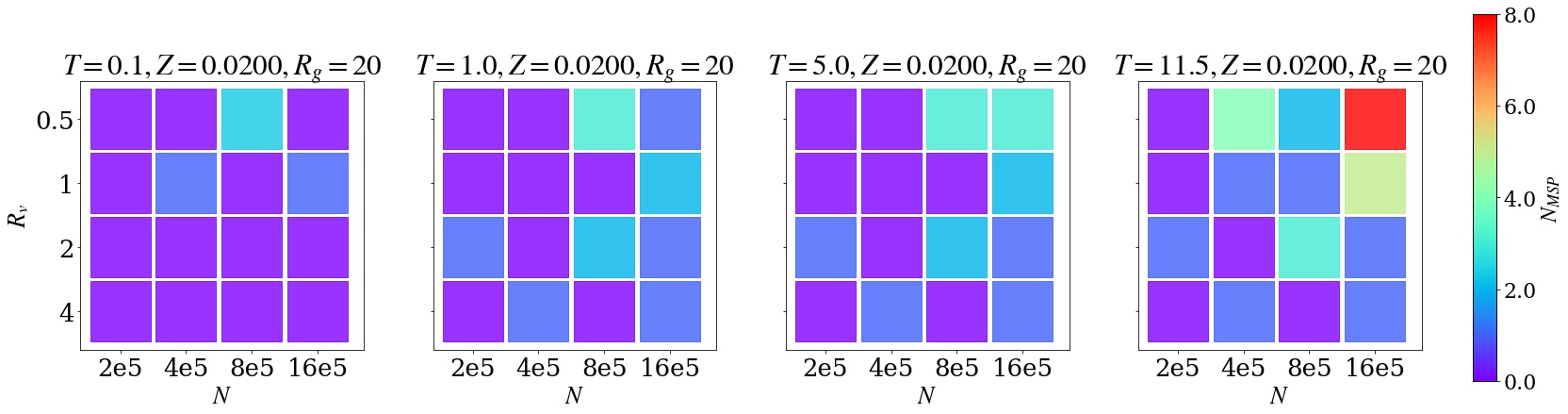}
    \caption{Same as Figure~\ref{fig:z0.002_rg8} but for models with a metallicity $Z=0.02$ and Galactocentric distance $R_g = 20~$kpc.}
    \label{fig:z0.02_rg20}
\end{figure*}

\newpage~\newpage~\newpage
\bibliographystyle{aasjournal}
\bibliography{MSPinGCE}

\begin{thebibliography}{}
\expandafter\ifx\csname natexlab\endcsname\relax\def\natexlab#1{#1}\fi
\providecommand{\url}[1]{\href{#1}{#1}}
\providecommand{\dodoi}[1]{doi:~\href{http://doi.org/#1}{\nolinkurl{#1}}}
\providecommand{\doeprint}[1]{\href{http://ascl.net/#1}{\nolinkurl{http://ascl.net/#1}}}
\providecommand{\doarXiv}[1]{\href{https://arxiv.org/abs/#1}{\nolinkurl{https://arxiv.org/abs/#1}}}

\bibitem[{{Abazajian}(2011)}]{Abazajian_2011}
{Abazajian}, K.~N. 2011, \jcap, 2011, 010,
  \dodoi{10.1088/1475-7516/2011/03/010}

\bibitem[{{Abazajian} \& {Kaplinghat}(2012)}]{Abazajian+2012}
{Abazajian}, K.~N., \& {Kaplinghat}, M. 2012, \prd, 86, 083511,
  \dodoi{10.1103/PhysRevD.86.083511}

\bibitem[{{Abbate} {et~al.}(2018){Abbate}, {Mastrobuono-Battisti}, {Colpi},
  {Possenti}, {Sippel}, \& {Dotti}}]{Abbate+2018}
{Abbate}, F., {Mastrobuono-Battisti}, A., {Colpi}, M., {et~al.} 2018, \mnras,
  473, 927, \dodoi{10.1093/mnras/stx2364}

\bibitem[{{Abbott} {et~al.}(2017){Abbott}, {Abbott}, {Abbott}, {Acernese},
  {Ackley}, {Adams}, {Adams}, {Addesso}, {Adhikari}, {Adya}, {Affeldt},
  {Afrough}, {Agarwal}, {Agathos}, {Agatsuma}, {Aggarwal}, {Aguiar}, {Aiello},
  {Ain}, {Ajith}, {Allen}, {Allen}, {Allocca}, {Altin}, {Amato}, {Ananyeva},
  {Anderson}, {Anderson}, {Angelova}, {Antier}, {Appert}, {Arai}, {Araya},
  {Areeda}, {Arnaud}, {Arun}, {Ascenzi}, {Ashton}, {Ast}, {Aston}, {Astone},
  {Atallah}, {Aufmuth}, {Aulbert}, {AultONeal}, {Austin}, {Avila-Alvarez},
  {Babak}, {Bacon}, {Bader}, {Bae}, {Bailes}, {Baker}, {Baldaccini},
  {Ballardin}, {Ballmer}, {Banagiri}, {Barayoga}, {Barclay}, {Barish},
  {Barker}, {Barkett}, {Barone}, {Barr}, {Barsotti}, {Barsuglia}, {Barta},
  {Barthelmy}, {Bartlett}, {Bartos}, {Bassiri}, {Basti}, {Batch}, {Bawaj},
  {Bayley}, {Bazzan}, {B{\'e}csy}, {Beer}, {Bejger}, {Belahcene}, {Bell},
  {Berger}, {Bergmann}, {Bernuzzi}, {Bero}, {Berry}, {Bersanetti}, {Bertolini},
  {Betzwieser}, {Bhagwat}, {Bhandare}, {Bilenko}, {Billingsley}, {Billman},
  {Birch}, {Birney}, {Birnholtz}, {Biscans}, {Biscoveanu}, {Bisht}, {Bitossi},
  {Biwer}, {Bizouard}, {Blackburn}, {Blackman}, {Blair}, {Blair}, {Blair},
  {Bloemen}, {Bock}, {Bode}, {Boer}, {Bogaert}, {Bohe}, {Bondu}, {Bonilla},
  {Bonnand}, {Boom}, {Bork}, {Boschi}, {Bose}, {Bossie}, {Bouffanais}, {Bozzi},
  {Bradaschia}, {Brady}, {Branchesi}, {Brau}, {Briant}, {Brillet}, {Brinkmann},
  {Brisson}, {Brockill}, {Broida}, {Brooks}, {Brown}, {Brown}, {Brunett},
  {Buchanan}, {Buikema}, {Bulik}, {Bulten}, {Buonanno}, {Buskulic}, {Buy},
  {Byer}, {Cabero}, {Cadonati}, {Cagnoli}, {Cahillane}, {Calder{\'o}n
  Bustillo}, {Callister}, {Calloni}, {Camp}, {Canepa}, {Canizares}, {Cannon},
  {Cao}, {Cao}, {Capano}, {Capocasa}, {Carbognani}, {Caride}, {Carney},
  {Carullo}, {Casanueva Diaz}, {Casentini}, {Caudill}, {Cavagli{\`a}},
  {Cavalier}, {Cavalieri}, {Cella}, {Cepeda}, {Cerd{\'a}-Dur{\'a}n},
  {Cerretani}, {Cesarini}, {Chamberlin}, {Chan}, {Chao}, {Charlton}, {Chase},
  {Chassande-Mottin}, {Chatterjee}, {Chatziioannou}, {Cheeseboro}, {Chen},
  {Chen}, {Chen}, {Cheng}, {Chia}, {Chincarini}, {Chiummo}, {Chmiel}, {Cho},
  {Cho}, {Chow}, {Christensen}, {Chu}, {Chua}, {Chua}, {Chung}, {Chung},
  {Ciani}, {Ciolfi}, {Cirelli}, {Cirone}, {Clara}, {Clark}, {Clearwater},
  {Cleva}, {Cocchieri}, {Coccia}, {Cohadon}, {Cohen}, {Colla}, {Collette},
  {Cominsky}, {Constancio}, {Conti}, {Cooper}, {Corban}, {Corbitt},
  {Cordero-Carri{\'o}n}, {Corley}, {Cornish}, {Corsi}, {Cortese}, {Costa},
  {Coughlin}, {Coughlin}, {Coulon}, {Countryman}, {Couvares}, {Covas}, {Cowan},
  {Coward}, {Cowart}, {Coyne}, {Coyne}, {Creighton}, {Creighton}, {Cripe},
  {Crowder}, {Cullen}, {Cumming}, {Cunningham}, {Cuoco}, {Dal Canton},
  {D{\'a}lya}, {Danilishin}, {D'Antonio}, {Danzmann}, {Dasgupta}, {Da Silva
  Costa}, {Dattilo}, {Dave}, {Davier}, {Davis}, {Daw}, {Day}, {De}, {DeBra},
  {Degallaix}, {De Laurentis}, {Del{\'e}glise}, {Del Pozzo}, {Demos}, {Denker},
  {Dent}, {De Pietri}, {Dergachev}, {De Rosa}, {DeRosa}, {De Rossi}, {DeSalvo},
  {de Varona}, {Devenson}, {Dhurandhar}, {D{\'\i}az}, {Dietrich}, {Di Fiore},
  {Di Giovanni}, {Di Girolamo}, {Di Lieto}, {Di Pace}, {Di Palma}, {Di Renzo},
  {Doctor}, {Dolique}, {Donovan}, {Dooley}, {Doravari}, {Dorrington},
  {Douglas}, {Dovale {\'A}lvarez}, {Downes}, {Drago}, {Dreissigacker},
  {Driggers}, {Du}, {Ducrot}, {Dudi}, {Dupej}, {Dwyer}, {Edo}, {Edwards},
  {Effler}, {Eggenstein}, {Ehrens}, {Eichholz}, {Eikenberry}, {Eisenstein},
  {Essick}, {Estevez}, {Etienne}, {Etzel}, {Evans}, {Evans}, {Factourovich},
  {Fafone}, {Fair}, {Fairhurst}, {Fan}, {Farinon}, {Farr}, {Farr},
  {Fauchon-Jones}, {Favata}, {Fays}, {Fee}, {Fehrmann}, {Feicht}, {Fejer},
  {Fernandez-Galiana}, {Ferrante}, {Ferreira}, {Ferrini}, {Fidecaro},
  {Finstad}, {Fiori}, {Fiorucci}, {Fishbach}, {Fisher}, {Fitz-Axen},
  {Flaminio}, {Fletcher}, {Fong}, {Font}, {Forsyth}, {Forsyth}, {Fournier},
  {Frasca}, {Frasconi}, {Frei}, {Freise}, {Frey}, {Frey}, {Fries}, {Fritschel},
  {Frolov}, {Fulda}, {Fyffe}, {Gabbard}, {Gadre}, {Gaebel}, {Gair},
  {Gammaitoni}, {Ganija}, {Gaonkar}, {Garcia-Quiros}, {Garufi}, {Gateley},
  {Gaudio}, {Gaur}, {Gayathri}, {Gehrels}, {Gemme}, {Genin}, {Gennai},
  {George}, {George}, {Gergely}, {Germain}, {Ghonge}, {Ghosh}, {Ghosh},
  {Ghosh}, {Giaime}, {Giardina}, {Giazotto}, {Gill}, {Glover}, {Goetz},
  {Goetz}, {Gomes}, {Goncharov}, {Gonz{\'a}lez}, {Gonzalez Castro},
  {Gopakumar}, {Gorodetsky}, {Gossan}, {Gosselin}, {Gouaty}, {Grado}, {Graef},
  {Granata}, {Grant}, {Gras}, {Gray}, {Greco}, {Green}, {Gretarsson}, {Groot},
  {Grote}, {Grunewald}, {Gruning}, {Guidi}, {Guo}, {Gupta}, {Gupta}, {Gushwa},
  {Gustafson}, {Gustafson}, {Halim}, {Hall}, {Hall}, {Hamilton}, {Hammond},
  {Haney}, {Hanke}, {Hanks}, {Hanna}, {Hannam}, {Hannuksela}, {Hanson},
  {Hardwick}, {Harms}, {Harry}, {Harry}, {Hart}, {Haster}, {Haughian}, {Healy},
  {Heidmann}, {Heintze}, {Heitmann}, {Hello}, {Hemming}, {Hendry}, {Heng},
  {Hennig}, {Heptonstall}, {Heurs}, {Hild}, {Hinderer}, {Ho}, {Hoak}, {Hofman},
  {Holt}, {Holz}, {Hopkins}, {Horst}, {Hough}, {Houston}, {Howell}, {Hreibi},
  {Hu}, {Huerta}, {Huet}, {Hughey}, {Husa}, {Huttner}, {Huynh-Dinh}, {Indik},
  {Inta}, {Intini}, {Isa}, {Isac}, {Isi}, {Iyer}, {Izumi}, {Jacqmin}, {Jani},
  {Jaranowski}, {Jawahar}, {Jim{\'e}nez-Forteza}, {Johnson},
  {Johnson-McDaniel}, {Jones}, {Jones}, {Jonker}, {Ju}, {Junker}, {Kalaghatgi},
  {Kalogera}, {Kamai}, {Kandhasamy}, {Kang}, {Kanner}, {Kapadia}, {Karki},
  {Karvinen}, {Kasprzack}, {Kastaun}, {Katolik}, {Katsavounidis}, {Katzman},
  {Kaufer}, {Kawabe}, {K{\'e}f{\'e}lian}, {Keitel}, {Kemball}, {Kennedy},
  {Kent}, {Key}, {Khalili}, {Khan}, {Khan}, {Khan}, {Khazanov}, {Kijbunchoo},
  {Kim}, {Kim}, {Kim}, {Kim}, {Kim}, {Kim}, {Kimbrell}, {King}, {King},
  {Kinley-Hanlon}, {Kirchhoff}, {Kissel}, {Kleybolte}, {Klimenko}, {Knowles},
  {Koch}, {Koehlenbeck}, {Koley}, {Kondrashov}, {Kontos}, {Korobko}, {Korth},
  {Kowalska}, {Kozak}, {Kr{\"a}mer}, {Kringel}, {Krishnan}, {Kr{\'o}lak},
  {Kuehn}, {Kumar}, {Kumar}, {Kumar}, {Kuo}, {Kutynia}, {Kwang}, {Lackey},
  {Lai}, {Landry}, {Lang}, {Lange}, {Lantz}, {Lanza}, {Larson},
  {Lartaux-Vollard}, {Lasky}, {Laxen}, {Lazzarini}, {Lazzaro}, {Leaci},
  {Leavey}, {Lee}, {Lee}, {Lee}, {Lee}, {Lee}, {Lehmann}, {Lenon}, {Leon},
  {Leonardi}, {Leroy}, {Letendre}, {Levin}, {Li}, {Linker}, {Littenberg},
  {Liu}, {Liu}, {Lo}, {Lockerbie}, {London}, {Lord}, {Lorenzini}, {Loriette},
  {Lormand}, {Losurdo}, {Lough}, {Lousto}, {Lovelace}, {L{\"u}ck}, {Lumaca},
  {Lundgren}, {Lynch}, {Ma}, {Macas}, {Macfoy}, {Machenschalk}, {MacInnis},
  {Macleod}, {Maga{\~n}a Hernandez}, {Maga{\~n}a-Sandoval}, {Maga{\~n}a
  Zertuche}, {Magee}, {Majorana}, {Maksimovic}, {Man}, {Mandic}, {Mangano},
  {Mansell}, {Manske}, {Mantovani}, {Marchesoni}, {Marion}, {M{\'a}rka},
  {M{\'a}rka}, {Markakis}, {Markosyan}, {Markowitz}, {Maros}, {Marquina},
  {Marsh}, {Martelli}, {Martellini}, {Martin}, {Martin}, {Martynov}, {Marx},
  {Mason}, {Massera}, {Masserot}, {Massinger}, {Masso-Reid}, {Mastrogiovanni},
  {Matas}, {Matichard}, {Matone}, {Mavalvala}, {Mazumder}, {McCarthy},
  {McClelland}, {McCormick}, {McCuller}, {McGuire}, {McIntyre}, {McIver},
  {McManus}, {McNeill}, {McRae}, {McWilliams}, {Meacher}, {Meadors}, {Mehmet},
  {Meidam}, {Mejuto-Villa}, {Melatos}, {Mendell}, {Mercer}, {Merilh},
  {Merzougui}, {Meshkov}, {Messenger}, {Messick}, {Metzdorff}, {Meyers},
  {Miao}, {Michel}, {Middleton}, {Mikhailov}, {Milano}, {Miller}, {Miller},
  {Miller}, {Millhouse}, {Milovich-Goff}, {Minazzoli}, {Minenkov}, {Ming},
  {Mishra}, {Mitra}, {Mitrofanov}, {Mitselmakher}, {Mittleman}, {Moffa},
  {Moggi}, {Mogushi}, {Mohan}, {Mohapatra}, {Molina}, {Montani}, {Moore},
  {Moraru}, {Moreno}, {Morisaki}, {Morriss}, {Mours}, {Mow-Lowry}, {Mueller},
  {Muir}, {Mukherjee}, {Mukherjee}, {Mukherjee}, {Mukund}, {Mullavey}, {Munch},
  {Mu{\~n}iz}, {Muratore}, {Murray}, {Nagar}, {Napier}, {Nardecchia},
  {Naticchioni}, {Nayak}, {Neilson}, {Nelemans}, {Nelson}, {Nery}, {Neunzert},
  {Nevin}, {Newport}, {Newton}, {Ng}, {Nguyen}, {Nguyen}, {Nichols}, {Nielsen},
  {Nissanke}, {Nitz}, {Noack}, {Nocera}, {Nolting}, {North}, {Nuttall},
  {Oberling}, {O'Dea}, {Ogin}, {Oh}, {Oh}, {Ohme}, {Okada}, {Oliver},
  {Oppermann}, {Oram}, {O'Reilly}, {Ormiston}, {Ortega}, {O'Shaughnessy},
  {Ossokine}, {Ottaway}, {Overmier}, {Owen}, {Pace}, {Page}, {Page}, {Pai},
  {Pai}, {Palamos}, {Palashov}, {Palomba}, {Pal-Singh}, {Pan}, {Pan}, {Pang},
  {Pang}, {Pankow}, {Pannarale}, {Pant}, {Paoletti}, {Paoli}, {Papa}, {Parida},
  {Parker}, {Pascucci}, {Pasqualetti}, {Passaquieti}, {Passuello}, {Patil},
  {Patricelli}, {Pearlstone}, {Pedraza}, {Pedurand}, {Pekowsky}, {Pele},
  {Penn}, {Perez}, {Perreca}, {Perri}, {Pfeiffer}, {Phelps}, {Piccinni},
  {Pichot}, {Piergiovanni}, {Pierro}, {Pillant}, {Pinard}, {Pinto}, {Pirello},
  {Pitkin}, {Poe}, {Poggiani}, {Popolizio}, {Porter}, {Post}, {Powell},
  {Prasad}, {Pratt}, {Pratten}, {Predoi}, {Prestegard}, {Prijatelj},
  {Principe}, {Privitera}, {Prix}, {Prodi}, {Prokhorov}, {Puncken}, {Punturo},
  {Puppo}, {P{\"u}rrer}, {Qi}, {Quetschke}, {Quintero}, {Quitzow-James},
  {Raab}, {Rabeling}, {Radkins}, {Raffai}, {Raja}, {Rajan}, {Rajbhandari},
  {Rakhmanov}, {Ramirez}, {Ramos-Buades}, {Rapagnani}, {Raymond}, {Razzano},
  {Read}, {Regimbau}, {Rei}, {Reid}, {Reitze}, {Ren}, {Reyes}, {Ricci},
  {Ricker}, {Rieger}, {Riles}, {Rizzo}, {Robertson}, {Robie}, {Robinet},
  {Rocchi}, {Rolland}, {Rollins}, {Roma}, {Romano}, {Romano}, {Romel}, {Romie},
  {Rosi{\'n}ska}, {Ross}, {Rowan}, {R{\"u}diger}, {Ruggi}, {Rutins}, {Ryan},
  {Sachdev}, {Sadecki}, {Sadeghian}, {Sakellariadou}, {Salconi}, {Saleem},
  {Salemi}, {Samajdar}, {Sammut}, {Sampson}, {Sanchez}, {Sanchez},
  {Sanchis-Gual}, {Sandberg}, {Sanders}, {Sassolas}, {Sathyaprakash},
  {Saulson}, {Sauter}, {Savage}, {Sawadsky}, {Schale}, {Scheel}, {Scheuer},
  {Schmidt}, {Schmidt}, {Schnabel}, {Schofield}, {Sch{\"o}nbeck}, {Schreiber},
  {Schuette}, {Schulte}, {Schutz}, {Schwalbe}, {Scott}, {Scott}, {Seidel},
  {Sellers}, {Sengupta}, {Sentenac}, {Sequino}, {Sergeev}, {Shaddock},
  {Shaffer}, {Shah}, {Shahriar}, {Shaner}, {Shao}, {Shapiro}, {Shawhan},
  {Sheperd}, {Shoemaker}, {Shoemaker}, {Siellez}, {Siemens}, {Sieniawska},
  {Sigg}, {Silva}, {Singer}, {Singh}, {Singhal}, {Sintes}, {Slagmolen},
  {Smith}, {Smith}, {Smith}, {Somala}, {Son}, {Sonnenberg}, {Sorazu},
  {Sorrentino}, {Souradeep}, {Spencer}, {Srivastava}, {Staats}, {Staley},
  {Steinke}, {Steinlechner}, {Steinlechner}, {Steinmeyer}, {Stevenson},
  {Stone}, {Stops}, {Strain}, {Stratta}, {Strigin}, {Strunk}, {Sturani},
  {Stuver}, {Summerscales}, {Sun}, {Sunil}, {Suresh}, {Sutton}, {Swinkels},
  {Szczepa{\'n}czyk}, {Tacca}, {Tait}, {Talbot}, {Talukder}, {Tanner},
  {T{\'a}pai}, {Taracchini}, {Tasson}, {Taylor}, {Taylor}, {Tewari}, {Theeg},
  {Thies}, {Thomas}, {Thomas}, {Thomas}, {Thorne}, {Thorne}, {Thrane},
  {Tiwari}, {Tiwari}, {Tokmakov}, {Toland}, {Tonelli}, {Tornasi},
  {Torres-Forn{\'e}}, {Torrie}, {T{\"o}yr{\"a}}, {Travasso}, {Traylor},
  {Trinastic}, {Tringali}, {Trozzo}, {Tsang}, {Tse}, {Tso}, {Tsukada}, {Tsuna},
  {Tuyenbayev}, {Ueno}, {Ugolini}, {Unnikrishnan}, {Urban}, {Usman},
  {Vahlbruch}, {Vajente}, {Valdes}, {Vallisneri}, {van Bakel}, {van Beuzekom},
  {van den Brand}, {Van Den Broeck}, {Vander-Hyde}, {van der Schaaf}, {van
  Heijningen}, {van Veggel}, {Vardaro}, {Varma}, {Vass}, {Vas{\'u}th},
  {Vecchio}, {Vedovato}, {Veitch}, {Veitch}, {Venkateswara}, {Venugopalan},
  {Verkindt}, {Vetrano}, {Vicer{\'e}}, {Viets}, {Vinciguerra}, {Vine}, {Vinet},
  {Vitale}, {Vo}, {Vocca}, {Vorvick}, {Vyatchanin}, {Wade}, {Wade}, {Wade},
  {Walet}, {Walker}, {Wallace}, {Walsh}, {Wang}, {Wang}, {Wang}, {Wang},
  {Wang}, {Ward}, {Warner}, {Was}, {Watchi}, {Weaver}, {Wei}, {Weinert},
  {Weinstein}, {Weiss}, {Wen}, {Wessel}, {We{\ss}els}, {Westerweck},
  {Westphal}, {Wette}, {Whelan}, {Whitcomb}, {Whiting}, {Whittle}, {Wilken},
  {Williams}, {Williams}, {Williamson}, {Willis}, {Willke}, {Wimmer},
  {Winkler}, {Wipf}, {Wittel}, {Woan}, {Woehler}, {Wofford}, {Wong}, {Worden},
  {Wright}, {Wu}, {Wysocki}, {Xiao}, {Yamamoto}, {Yancey}, {Yang}, {Yap},
  {Yazback}, {Yu}, {Yu}, {Yvert}, {Zadro{\.Z}ny}, {Zanolin}, {Zelenova},
  {Zendri}, {Zevin}, {Zhang}, {Zhang}, {Zhang}, {Zhang}, {Zhao}, {Zhou},
  {Zhou}, {Zhu}, {Zhu}, {Zimmerman}, {Zucker}, {Zweizig}, {LIGO Scientific
  Collaboration}, \& {Virgo Collaboration}}]{GW170817}
{Abbott}, B.~P., {Abbott}, R., {Abbott}, T.~D., {et~al.} 2017, \prl, 119,
  161101, \dodoi{10.1103/PhysRevLett.119.161101}

\bibitem[{{Abdo} {et~al.}(2009){Abdo}, {Ackermann}, {Ajello}, {Atwood},
  {Axelsson}, {Baldini}, {Ballet}, {Barbiellini}, {Bastieri}, {Baughman},
  {Bechtol}, {Bellazzini}, {Berenji}, {Blandford}, {Bloom}, {Bonamente},
  {Borgland}, {Bregeon}, {Brez}, {Brigida}, {Bruel}, {Burnett}, {Caliandro},
  {Cameron}, {Caraveo}, {Casandjian}, {Cecchi}, {{\c{C}}elik}, {Charles},
  {Chaty}, {Chekhtman}, {Cheung}, {Chiang}, {Ciprini}, {Claus}, {Cohen-Tanugi},
  {Conrad}, {Cutini}, {Dermer}, {de Palma}, {Digel}, {Dormody}, {do Couto e
  Silva}, {Drell}, {Dubois}, {Dumora}, {Farnier}, {Favuzzi}, {Fegan}, {Focke},
  {Frailis}, {Fukazawa}, {Fusco}, {Gargano}, {Gasparrini}, {Gehrels},
  {Germani}, {Giebels}, {Giglietto}, {Giordano}, {Glanzman}, {Godfrey},
  {Grenier}, {Grove}, {Guillemot}, {Guiriec}, {Hanabata}, {Harding},
  {Hayashida}, {Hays}, {Horan}, {Hughes}, {J{\'o}hannesson}, {Johnson},
  {Johnson}, {Johnson}, {Johnson}, {Kamae}, {Katagiri}, {Kawai}, {Kerr},
  {Kn{\"o}dlseder}, {Kuehn}, {Kuss}, {Lande}, {Latronico}, {Lemoine-Goumard},
  {Longo}, {Loparco}, {Lott}, {Lovellette}, {Lubrano}, {Makeev}, {Mazziotta},
  {McConville}, {McEnery}, {Meurer}, {Michelson}, {Mitthumsiri}, {Mizuno},
  {Moiseev}, {Monte}, {Monzani}, {Morselli}, {Moskalenko}, {Murgia}, {Nolan},
  {Norris}, {Nuss}, {Ohsugi}, {Omodei}, {Orlando}, {Ormes}, {Paneque},
  {Panetta}, {Parent}, {Pelassa}, {Pepe}, {Pierbattista}, {Piron}, {Porter},
  {Rain{\`o}}, {Rando}, {Razzano}, {Rea}, {Reimer}, {Reimer}, {Reposeur},
  {Ritz}, {Rochester}, {Rodriguez}, {Romani}, {Roth}, {Ryde}, {Sadrozinski},
  {Sanchez}, {Sander}, {Saz Parkinson}, {Sgr{\`o}}, {Smith}, {Smith},
  {Spandre}, {Spinelli}, {Starck}, {Strickman}, {Suson}, {Tajima}, {Takahashi},
  {Tanaka}, {Thayer}, {Thayer}, {Thompson}, {Tibaldo}, {Torres}, {Tosti},
  {Tramacere}, {Uchiyama}, {Usher}, {Vasileiou}, {Vilchez}, {Vitale}, {Wang},
  {Webb}, {Winer}, {Wood}, {Ylinen}, \& {Ziegler}}]{Abdo+2009}
{Abdo}, A.~A., {Ackermann}, M., {Ajello}, M., {et~al.} 2009, Science, 325, 845,
  \dodoi{10.1126/science.1177023}

\bibitem[{{Abdo} {et~al.}(2010){Abdo}, {Ackermann}, {Ajello}, {Baldini},
  {Ballet}, {Barbiellini}, {Bastieri}, {Bellazzini}, {Blandford}, {Bloom},
  {Bonamente}, {Borgland}, {Bouvier}, {Brandt}, {Bregeon}, {Brigida}, {Bruel},
  {Buehler}, {Buson}, {Caliandro}, {Cameron}, {Caraveo}, {Carrigan},
  {Casandjian}, {Charles}, {Chaty}, {Chekhtman}, {Cheung}, {Chiang}, {Ciprini},
  {Claus}, {Cohen-Tanugi}, {Conrad}, {Decesar}, {Dermer}, {de Palma}, {Digel},
  {Silva}, {Drell}, {Dubois}, {Dumora}, {Favuzzi}, {Fortin}, {Frailis},
  {Fukazawa}, {Fusco}, {Gargano}, {Gasparrini}, {Gehrels}, {Germani},
  {Giglietto}, {Giordano}, {Glanzman}, {Godfrey}, {Grenier}, {Grondin},
  {Grove}, {Guillemot}, {Guiriec}, {Hadasch}, {Harding}, {Hays}, {Jean},
  {J{\'o}hannesson}, {Johnson}, {Johnson}, {Kamae}, {Katagiri}, {Kataoka},
  {Kerr}, {Kn{\"o}dlseder}, {Kuss}, {Lande}, {Latronico}, {Lee},
  {Lemoine-Goumard}, {Llena Garde}, {Longo}, {Loparco}, {Lovellette},
  {Lubrano}, {Makeev}, {Mazziotta}, {Michelson}, {Mitthumsiri}, {Mizuno},
  {Monte}, {Monzani}, {Morselli}, {Moskalenko}, {Murgia}, {Naumann-Godo},
  {Nolan}, {Norris}, {Nuss}, {Ohsugi}, {Omodei}, {Orlando}, {Ormes},
  {Pancrazi}, {Parent}, {Pepe}, {Pesce-Rollins}, {Piron}, {Porter},
  {Rain{\`o}}, {Rando}, {Reimer}, {Reimer}, {Reposeur}, {Ripken}, {Romani},
  {Roth}, {Sadrozinski}, {Saz Parkinson}, {Sgr{\`o}}, {Siskind}, {Smith},
  {Spinelli}, {Strickman}, {Suson}, {Takahashi}, {Takahashi}, {Tanaka},
  {Thayer}, {Thayer}, {Tibaldo}, {Torres}, {Tosti}, {Tramacere}, {Uchiyama},
  {Usher}, {Vasileiou}, {Venter}, {Vilchez}, {Vitale}, {Waite}, {Wang}, {Webb},
  {Winer}, {Yang}, {Ylinen}, {Ziegler}, \& {Fermi LAT
  Collaboration}}]{Abdo+2010}
---. 2010, \aap, 524, A75, \dodoi{10.1051/0004-6361/201014458}

\bibitem[{{Alpar} {et~al.}(1982){Alpar}, {Cheng}, {Ruderman}, \&
  {Shaham}}]{Alpar+1982}
{Alpar}, M.~A., {Cheng}, A.~F., {Ruderman}, M.~A., \& {Shaham}, J. 1982, \nat,
  300, 728, \dodoi{10.1038/300728a0}

\bibitem[{{Bagchi} {et~al.}(2011){Bagchi}, {Lorimer}, \&
  {Chennamangalam}}]{Bagchi+2011}
{Bagchi}, M., {Lorimer}, D.~R., \& {Chennamangalam}, J. 2011, \mnras, 418, 477,
  \dodoi{10.1111/j.1365-2966.2011.19498.x}

\bibitem[{{Baumgardt}(2017)}]{Baumgardt_2017}
{Baumgardt}, H. 2017, \mnras, 464, 2174, \dodoi{10.1093/mnras/stw2488}

\bibitem[{{Baumgardt} \& {Hilker}(2018)}]{Baumgardt_Hilker2018}
{Baumgardt}, H., \& {Hilker}, M. 2018, \mnras, 478, 1520,
  \dodoi{10.1093/mnras/sty1057}

\bibitem[{{Baumgardt} \& {Vasiliev}(2021)}]{Baumgardt+2021}
{Baumgardt}, H., \& {Vasiliev}, E. 2021, \mnras, 505, 5957,
  \dodoi{10.1093/mnras/stab1474}

\bibitem[{{Bhattacharya} \& {van den
  Heuvel}(1991)}]{Bhattacharya_vandenHeuvel_1991}
{Bhattacharya}, D., \& {van den Heuvel}, E.~P.~J. 1991, \physrep, 203, 1,
  \dodoi{10.1016/0370-1573(91)90064-S}

\bibitem[{{Binney} \& {Tremaine}(2008)}]{BT_galacticdynamics}
{Binney}, J., \& {Tremaine}, S. 2008, {Galactic Dynamics: Second Edition}

\bibitem[{{Brandt} \& {Kocsis}(2015)}]{Brandt_Kocsis2015}
{Brandt}, T.~D., \& {Kocsis}, B. 2015, \apj, 812, 15,
  \dodoi{10.1088/0004-637X/812/1/15}

\bibitem[{{Breivik} {et~al.}(2020){Breivik}, {Coughlin}, {Zevin}, {Rodriguez},
  {Kremer}, {Ye}, {Andrews}, {Kurkowski}, {Digman}, {Larson}, \&
  {Rasio}}]{cosmic}
{Breivik}, K., {Coughlin}, S., {Zevin}, M., {et~al.} 2020, \apj, 898, 71,
  \dodoi{10.3847/1538-4357/ab9d85}

\bibitem[{{Cadelano} {et~al.}(2020){Cadelano}, {Dalessandro}, {Webb},
  {Vesperini}, {Lattanzio}, {Beccari}, {Gomez}, \& {Monaco}}]{Cadelano+2020}
{Cadelano}, M., {Dalessandro}, E., {Webb}, J.~J., {et~al.} 2020, \mnras, 499,
  2390, \dodoi{10.1093/mnras/staa2759}

\bibitem[{{Calore} {et~al.}(2015){Calore}, {Cholis}, \&
  {Weniger}}]{Calore+2015a}
{Calore}, F., {Cholis}, I., \& {Weniger}, C. 2015, \jcap, 2015, 038,
  \dodoi{10.1088/1475-7516/2015/03/038}

\bibitem[{Chatterjee {et~al.}(2010)Chatterjee, Fregeau, Umbreit, \&
  Rasio}]{Chatterjee_2010}
Chatterjee, S., Fregeau, J.~M., Umbreit, S., \& Rasio, F.~A. 2010, \apj, 719,
  915.
\newblock \url{http://dx.doi.org/10.1088/0004-637X/719/1/915}

\bibitem[{Chatterjee {et~al.}(2013)Chatterjee, Umbreit, Fregeau, \&
  Rasio}]{Chatterjee_2013b}
Chatterjee, S., Umbreit, S., Fregeau, J.~M., \& Rasio, F.~A. 2013, \mnras, 429,
  2881.
\newblock \url{http://dx.doi.org/10.1093/mnras/sts464}

\bibitem[{{Cholis} {et~al.}(2015){Cholis}, {Hooper}, \& {Linden}}]{Cholis+2015}
{Cholis}, I., {Hooper}, D., \& {Linden}, T. 2015, \jcap, 2015, 043,
  \dodoi{10.1088/1475-7516/2015/06/043}

\bibitem[{{Clark}(1975)}]{Clark_1975}
{Clark}, G.~W. 1975, \apjl, 199, L143, \dodoi{10.1086/181869}

\bibitem[{{Daylan} {et~al.}(2016){Daylan}, {Finkbeiner}, {Hooper}, {Linden},
  {Portillo}, {Rodd}, \& {Slatyer}}]{Daylan+2016}
{Daylan}, T., {Finkbeiner}, D.~P., {Hooper}, D., {et~al.} 2016, Physics of the
  Dark Universe, 12, 1, \dodoi{10.1016/j.dark.2015.12.005}

\bibitem[{{Di Mauro}(2021)}]{DiMauro2021}
{Di Mauro}, M. 2021, \prd, 103, 063029, \dodoi{10.1103/PhysRevD.103.063029}

\bibitem[{{Dinsmore} \& {Slatyer}(2021)}]{Dinsmore_Slatyer_2021}
{Dinsmore}, J.~T., \& {Slatyer}, T.~R. 2021, arXiv e-prints, arXiv:2112.09699.
\newblock \doarXiv{2112.09699}

\bibitem[{{Dotter} {et~al.}(2011){Dotter}, {Sarajedini}, \&
  {Anderson}}]{Dotter+2011}
{Dotter}, A., {Sarajedini}, A., \& {Anderson}, J. 2011, \apj, 738, 74,
  \dodoi{10.1088/0004-637X/738/1/74}

\bibitem[{{Dotter} {et~al.}(2010){Dotter}, {Sarajedini}, {Anderson},
  {Aparicio}, {Bedin}, {Chaboyer}, {Majewski}, {Mar{\'\i}n-Franch}, {Milone},
  {Paust}, {Piotto}, {Reid}, {Rosenberg}, \& {Siegel}}]{Dotter+2010}
{Dotter}, A., {Sarajedini}, A., {Anderson}, J., {et~al.} 2010, \apj, 708, 698,
  \dodoi{10.1088/0004-637X/708/1/698}

\bibitem[{{Duquennoy} \& {Mayor}(1991)}]{Duquennoy_Mayor_1991}
{Duquennoy}, A., \& {Mayor}, M. 1991, \aap, 500, 337

\bibitem[{{Forbes} \& {Bridges}(2010)}]{Forbes_Bridges_2010}
{Forbes}, D.~A., \& {Bridges}, T. 2010, \mnras, 404, 1203,
  \dodoi{10.1111/j.1365-2966.2010.16373.x}

\bibitem[{{Fragione} {et~al.}(2018){Fragione}, {Antonini}, \&
  {Gnedin}}]{Fragione+2018gce}
{Fragione}, G., {Antonini}, F., \& {Gnedin}, O.~Y. 2018, \mnras, 475, 5313,
  \dodoi{10.1093/mnras/sty183}

\bibitem[{Fregeau {et~al.}(2004)Fregeau, Cheung, Portegies~Zwart, \&
  Rasio}]{fregeau2004stellar}
Fregeau, J.~M., Cheung, P., Portegies~Zwart, S., \& Rasio, F. 2004, \mnras,
  352, 1

\bibitem[{Fregeau {et~al.}(2003)Fregeau, Gurkan, Joshi, \&
  Rasio}]{Fregeau_2003}
Fregeau, J.~M., Gurkan, M.~A., Joshi, K.~J., \& Rasio, F.~A. 2003, \apj, 593,
  772.
\newblock \url{http://dx.doi.org/10.1086/376593}

\bibitem[{Fregeau \& Rasio(2007)}]{fregeau2007monte}
Fregeau, J.~M., \& Rasio, F.~A. 2007, \apj, 658, 1047

\bibitem[{{Fruchter} \& {Goss}(1990)}]{Fruchter_Goss_1990}
{Fruchter}, A.~S., \& {Goss}, W.~M. 1990, \apjl, 365, L63,
  \dodoi{10.1086/185889}

\bibitem[{{Gautam} {et~al.}(2022){Gautam}, {Crocker}, {Ferrario}, {Ruiter},
  {Ploeg}, {Gordon}, \& {Macias}}]{Gautam+2022}
{Gautam}, A., {Crocker}, R.~M., {Ferrario}, L., {et~al.} 2022, Nature
  Astronomy, \dodoi{10.1038/s41550-022-01658-3}

\bibitem[{{Gieles} \& {Baumgardt}(2008)}]{Gieles_Baumgardt_2008}
{Gieles}, M., \& {Baumgardt}, H. 2008, \mnras, 389, L28,
  \dodoi{10.1111/j.1745-3933.2008.00515.x}

\bibitem[{{Gnedin} {et~al.}(2014){Gnedin}, {Ostriker}, \&
  {Tremaine}}]{Gnedin+2014}
{Gnedin}, O.~Y., {Ostriker}, J.~P., \& {Tremaine}, S. 2014, \apj, 785, 71,
  \dodoi{10.1088/0004-637X/785/1/71}

\bibitem[{{Gonthier} {et~al.}(2018){Gonthier}, {Harding}, {Ferrara},
  {Frederick}, {Mohr}, \& {Koh}}]{Gonthier+2018}
{Gonthier}, P.~L., {Harding}, A.~K., {Ferrara}, E.~C., {et~al.} 2018, \apj,
  863, 199, \dodoi{10.3847/1538-4357/aad08d}

\bibitem[{{Goodenough} \& {Hooper}(2009)}]{Goodenough_Hooper_2009}
{Goodenough}, L., \& {Hooper}, D. 2009, arXiv e-prints, arXiv:0910.2998.
\newblock \doarXiv{0910.2998}

\bibitem[{{Haggard} {et~al.}(2017){Haggard}, {Heinke}, {Hooper}, \&
  {Linden}}]{Haggard+2017}
{Haggard}, D., {Heinke}, C., {Hooper}, D., \& {Linden}, T. 2017, \jcap, 2017,
  056, \dodoi{10.1088/1475-7516/2017/05/056}

\bibitem[{{Haghi} {et~al.}(2017){Haghi}, {Khalaj}, {Hasani Zonoozi}, \&
  {Kroupa}}]{Haghi+2017}
{Haghi}, H., {Khalaj}, P., {Hasani Zonoozi}, A., \& {Kroupa}, P. 2017, \apj,
  839, 60, \dodoi{10.3847/1538-4357/aa6719}

\bibitem[{{Hansen} \& {Kawaler}(1994)}]{Hansen_Kawaler_1994}
{Hansen}, C.~J., \& {Kawaler}, S.~D. 1994, {Stellar Interiors. Physical
  Principles, Structure, and Evolution.}, \dodoi{10.1007/978-1-4419-9110-2}

\bibitem[{{Harris}(1996)}]{Harris_1996}
{Harris}, W.~E. 1996, \aj, 112, 1487, \dodoi{10.1086/118116}

\bibitem[{{Heggie}(1975)}]{Heggie1975}
{Heggie}, D.~C. 1975, \mnras, 173, 729, \dodoi{10.1093/mnras/173.3.729}

\bibitem[{{Heinke} {et~al.}(2005){Heinke}, {Grindlay}, {Edmonds}, {Cohn},
  {Lugger}, {Camilo}, {Bogdanov}, \& {Freire}}]{Heinke+2005}
{Heinke}, C.~O., {Grindlay}, J.~E., {Edmonds}, P.~D., {et~al.} 2005, \apj, 625,
  796, \dodoi{10.1086/429899}

\bibitem[{{H{\'e}nault-Brunet} {et~al.}(2020){H{\'e}nault-Brunet}, {Gieles},
  {Strader}, {Peuten}, {Balbinot}, \& {Douglas}}]{Henault-Brunet+2020}
{H{\'e}nault-Brunet}, V., {Gieles}, M., {Strader}, J., {et~al.} 2020, \mnras,
  491, 113, \dodoi{10.1093/mnras/stz2995}

\bibitem[{H{\'e}non(1971{\natexlab{a}})}]{henon1971monte}
H{\'e}non, M. 1971{\natexlab{a}}, in International Astronomical Union
  Colloquium, Vol.~10, Cambridge University Press, 151--167

\bibitem[{H{\'e}non(1971{\natexlab{b}})}]{henon1971montecluster}
H{\'e}non, M. 1971{\natexlab{b}}, \apss, 13, 284

\bibitem[{{Hewish} {et~al.}(1968){Hewish}, {Bell}, {Pilkington}, {Scott}, \&
  {Collins}}]{Hewish+1968}
{Hewish}, A., {Bell}, S.~J., {Pilkington}, J.~D.~H., {Scott}, P.~F., \&
  {Collins}, R.~A. 1968, \nat, 217, 709, \dodoi{10.1038/217709a0}

\bibitem[{{Hobbs} {et~al.}(2005){Hobbs}, {Lorimer}, {Lyne}, \&
  {Kramer}}]{Hobbs+2005}
{Hobbs}, G., {Lorimer}, D.~R., {Lyne}, A.~G., \& {Kramer}, M. 2005, \mnras,
  360, 974, \dodoi{10.1111/j.1365-2966.2005.09087.x}

\bibitem[{{Hooper} \& {Goodenough}(2011)}]{Hooper_Goodenough_2011}
{Hooper}, D., \& {Goodenough}, L. 2011, Physics Letters B, 697, 412,
  \dodoi{10.1016/j.physletb.2011.02.029}

\bibitem[{{Hooper} \& {Linden}(2011)}]{Hooper_Linden_2011}
{Hooper}, D., \& {Linden}, T. 2011, \prd, 84, 123005,
  \dodoi{10.1103/PhysRevD.84.123005}

\bibitem[{{Hooper} \& {Linden}(2016)}]{Hooper_Linden_2016}
---. 2016, \jcap, 2016, 018, \dodoi{10.1088/1475-7516/2016/08/018}

\bibitem[{{Hooper} \& {Mohlabeng}(2016)}]{Hooper_Mohlabeng2016}
{Hooper}, D., \& {Mohlabeng}, G. 2016, \jcap, 2016, 049,
  \dodoi{10.1088/1475-7516/2016/03/049}

\bibitem[{{Hooper} \& {Slatyer}(2013)}]{Hooper_Slatype_2013}
{Hooper}, D., \& {Slatyer}, T.~R. 2013, Physics of the Dark Universe, 2, 118,
  \dodoi{10.1016/j.dark.2013.06.003}

\bibitem[{{Horiuchi} {et~al.}(2016){Horiuchi}, {Kaplinghat}, \&
  {Kwa}}]{Horiuchi+2016}
{Horiuchi}, S., {Kaplinghat}, M., \& {Kwa}, A. 2016, \jcap, 2016, 053,
  \dodoi{10.1088/1475-7516/2016/11/053}

\bibitem[{{Hui} {et~al.}(2010){Hui}, {Cheng}, \& {Taam}}]{Hui+2010}
{Hui}, C.~Y., {Cheng}, K.~S., \& {Taam}, R.~E. 2010, \apj, 714, 1149,
  \dodoi{10.1088/0004-637X/714/2/1149}

\bibitem[{Hurley {et~al.}(2000)Hurley, Pols, \& Tout}]{hurley2000comprehensive}
Hurley, J.~R., Pols, O.~R., \& Tout, C.~A. 2000, \mnras, 315, 543

\bibitem[{Hurley {et~al.}(2002)Hurley, Tout, \& Pols}]{hurley2002evolution}
Hurley, J.~R., Tout, C.~A., \& Pols, O.~R. 2002, \mnras, 329, 897

\bibitem[{{Ivanova} {et~al.}(2008){Ivanova}, {Heinke}, {Rasio}, {Belczynski},
  \& {Fregeau}}]{Ivanova+2008}
{Ivanova}, N., {Heinke}, C.~O., {Rasio}, F.~A., {Belczynski}, K., \& {Fregeau},
  J.~M. 2008, \mnras, 386, 553, \dodoi{10.1111/j.1365-2966.2008.13064.x}

\bibitem[{Joshi {et~al.}(2001)Joshi, Nave, \& Rasio}]{Joshi_2001}
Joshi, K.~J., Nave, C.~P., \& Rasio, F.~A. 2001, \apj, 550, 691.
\newblock \url{http://dx.doi.org/10.1086/319771}

\bibitem[{Joshi {et~al.}(2000)Joshi, Rasio, \& Portegies~Zwart}]{Joshi_2000}
Joshi, K.~J., Rasio, F.~A., \& Portegies~Zwart, S. 2000, \apj, 540, 969.
\newblock \url{http://dx.doi.org/10.1086/309350}

\bibitem[{{Kaspi}(2010)}]{Kaspi_2010}
{Kaspi}, V.~M. 2010, Proceedings of the National Academy of Science, 107, 7147,
  \dodoi{10.1073/pnas.1000812107}

\bibitem[{{Katz}(1975)}]{Katz_1975}
{Katz}, J.~I. 1975, \nat, 253, 698, \dodoi{10.1038/253698a0}

\bibitem[{{Kiel} {et~al.}(2008){Kiel}, {Hurley}, {Bailes}, \&
  {Murray}}]{Kiel+2008}
{Kiel}, P.~D., {Hurley}, J.~R., {Bailes}, M., \& {Murray}, J.~R. 2008, \mnras,
  388, 393, \dodoi{10.1111/j.1365-2966.2008.13402.x}

\bibitem[{{King}(1966)}]{King1966}
{King}, I.~R. 1966, \aj, 71, 64, \dodoi{10.1086/109857}

\bibitem[{{Kremer} {et~al.}(2022){Kremer}, {Ye}, {K{\i}ro{\u{g}}lu},
  {Lombardi}, {Ransom}, \& {Rasio}}]{Kremer_nstde_2022}
{Kremer}, K., {Ye}, C.~S., {K{\i}ro{\u{g}}lu}, F., {et~al.} 2022, arXiv
  e-prints, arXiv:2204.07169.
\newblock \doarXiv{2204.07169}

\bibitem[{{Kremer} {et~al.}(2020){Kremer}, {Ye}, {Rui}, {Weatherford},
  {Chatterjee}, {Fragione}, {Rodriguez}, {Spera}, \&
  {Rasio}}]{Kremer+2020catalog}
{Kremer}, K., {Ye}, C.~S., {Rui}, N.~Z., {et~al.} 2020, \apjs, 247, 48,
  \dodoi{10.3847/1538-4365/ab7919}

\bibitem[{Kroupa(2001)}]{Kroupa2001}
Kroupa, P. 2001, \mnras, 322, 231

\bibitem[{{Kruijssen} {et~al.}(2019){Kruijssen}, {Pfeffer}, {Reina-Campos},
  {Crain}, \& {Bastian}}]{Kruijssen+2019}
{Kruijssen}, J.~M.~D., {Pfeffer}, J.~L., {Reina-Campos}, M., {Crain}, R.~A., \&
  {Bastian}, N. 2019, \mnras, 486, 3180, \dodoi{10.1093/mnras/sty1609}

\bibitem[{{Kulkarni} {et~al.}(1990){Kulkarni}, {Narayan}, \&
  {Romani}}]{Kulkarni+1990}
{Kulkarni}, S.~R., {Narayan}, R., \& {Romani}, R.~W. 1990, \apj, 356, 174,
  \dodoi{10.1086/168828}

\bibitem[{{Launhardt} {et~al.}(2002){Launhardt}, {Zylka}, \&
  {Mezger}}]{Launhardt+2002}
{Launhardt}, R., {Zylka}, R., \& {Mezger}, P.~G. 2002, \aap, 384, 112,
  \dodoi{10.1051/0004-6361:20020017}

\bibitem[{{Lorimer}(2008)}]{Lorimer_2008}
{Lorimer}, D.~R. 2008, Living Reviews in Relativity, 11, 8,
  \dodoi{10.12942/lrr-2008-8}

\bibitem[{{Marks} {et~al.}(2012){Marks}, {Kroupa}, {Dabringhausen}, \&
  {Pawlowski}}]{Marks+2012}
{Marks}, M., {Kroupa}, P., {Dabringhausen}, J., \& {Pawlowski}, M.~S. 2012,
  \mnras, 422, 2246, \dodoi{10.1111/j.1365-2966.2012.20767.x}

\bibitem[{{Merritt}(2013)}]{Merritt_2013}
{Merritt}, D. 2013, {Dynamics and Evolution of Galactic Nuclei}

\bibitem[{{Morscher} {et~al.}(2015){Morscher}, {Pattabiraman}, {Rodriguez},
  {Rasio}, \& {Umbreit}}]{Morscher+2015}
{Morscher}, M., {Pattabiraman}, B., {Rodriguez}, C., {Rasio}, F.~A., \&
  {Umbreit}, S. 2015, \apj, 800, 9, \dodoi{10.1088/0004-637X/800/1/9}

\bibitem[{{Murgia}(2020)}]{Murgia_2020}
{Murgia}, S. 2020, Annual Review of Nuclear and Particle Science, 70, 455,
  \dodoi{10.1146/annurev-nucl-101916-123029}

\bibitem[{{Naiman} {et~al.}(2020){Naiman}, {Soares-Furtado}, \&
  {Ramirez-Ruiz}}]{Naiman+2020}
{Naiman}, J.~P., {Soares-Furtado}, M., \& {Ramirez-Ruiz}, E. 2020, \mnras, 491,
  4602, \dodoi{10.1093/mnras/stz3353}

\bibitem[{{Pan} {et~al.}(2021){Pan}, {Qian}, {Ma}, {Liu}, {Wang}, {Luo}, {Yan},
  {Ransom}, {Lorimer}, {Li}, \& {Jiang}}]{Pan+2021}
{Pan}, Z., {Qian}, L., {Ma}, X., {et~al.} 2021, \apjl, 915, L28,
  \dodoi{10.3847/2041-8213/ac0bbd}

\bibitem[{Pattabiraman {et~al.}(2013)Pattabiraman, Umbreit, Liao, Choudhary,
  Kalogera, Memik, \& Rasio}]{Pattabiraman_2013}
Pattabiraman, B., Umbreit, S., Liao, W.-k., {et~al.} 2013, \apjs, 204, 15.
\newblock \url{http://dx.doi.org/10.1088/0067-0049/204/2/15}

\bibitem[{{Petroff} {et~al.}(2022){Petroff}, {Hessels}, \&
  {Lorimer}}]{Petroff+2022}
{Petroff}, E., {Hessels}, J.~W.~T., \& {Lorimer}, D.~R. 2022, \aapr, 30, 2,
  \dodoi{10.1007/s00159-022-00139-w}

\bibitem[{{Ransom}(2008)}]{Ransom_2008}
{Ransom}, S.~M. 2008, in Dynamical Evolution of Dense Stellar Systems, ed.
  E.~{Vesperini}, M.~{Giersz}, \& A.~{Sills}, Vol. 246, 291--300,
  \dodoi{10.1017/S1743921308015810}

\bibitem[{{Ridolfi} {et~al.}(2021){Ridolfi}, {Gautam}, {Freire}, {Ransom},
  {Buchner}, {Possenti}, {Venkatraman Krishnan}, {Bailes}, {Kramer},
  {Stappers}, {Abbate}, {Barr}, {Burgay}, {Camilo}, {Corongiu}, {Jameson},
  {Padmanabh}, {Vleeschower}, {Champion}, {Chen}, {Geyer}, {Karastergiou},
  {Karuppusamy}, {Parthasarathy}, {Reardon}, {Serylak}, {Shannon}, \&
  {Spiewak}}]{Ridolfi+2021}
{Ridolfi}, A., {Gautam}, T., {Freire}, P.~C.~C., {et~al.} 2021, \mnras, 504,
  1407, \dodoi{10.1093/mnras/stab790}

\bibitem[{{Rodriguez} {et~al.}(2016){Rodriguez}, {Morscher}, {Wang},
  {Chatterjee}, {Rasio}, \& {Spurzem}}]{Rodriguez+2016million}
{Rodriguez}, C.~L., {Morscher}, M., {Wang}, L., {et~al.} 2016, \mnras, 463,
  2109, \dodoi{10.1093/mnras/stw2121}

\bibitem[{{Rodriguez} {et~al.}(2021){Rodriguez}, {Weatherford}, {Coughlin},
  {Seoane}, {Breivik}, {Chatterjee}, {Fragione}, {K{\i}ro{\u{g}}lu}, {Kremer},
  {Rui}, {Ye}, {Zevin}, \& {Rasio}}]{Rodriguez+2021CMC}
{Rodriguez}, C.~L., {Weatherford}, N.~C., {Coughlin}, S.~C., {et~al.} 2021,
  arXiv e-prints, arXiv:2106.02643.
\newblock \doarXiv{2106.02643}

\bibitem[{{Sch{\"o}del} {et~al.}(2014{\natexlab{a}}){Sch{\"o}del}, {Feldmeier},
  {Kunneriath}, {Stolovy}, {Neumayer}, {Amaro-Seoane}, \&
  {Nishiyama}}]{Schodel+2014a}
{Sch{\"o}del}, R., {Feldmeier}, A., {Kunneriath}, D., {et~al.}
  2014{\natexlab{a}}, \aap, 566, A47, \dodoi{10.1051/0004-6361/201423481}

\bibitem[{{Sch{\"o}del} {et~al.}(2014{\natexlab{b}}){Sch{\"o}del}, {Feldmeier},
  {Neumayer}, {Meyer}, \& {Yelda}}]{Schodel+2014b}
{Sch{\"o}del}, R., {Feldmeier}, A., {Neumayer}, N., {Meyer}, L., \& {Yelda}, S.
  2014{\natexlab{b}}, Classical and Quantum Gravity, 31, 244007,
  \dodoi{10.1088/0264-9381/31/24/244007}

\bibitem[{{Sigurdsson} \& {Phinney}(1995)}]{Sigurdsson_Phinney_1995}
{Sigurdsson}, S., \& {Phinney}, E.~S. 1995, \apjs, 99, 609,
  \dodoi{10.1086/192199}

\bibitem[{{Sollima} \& {Baumgardt}(2017)}]{Sollima_Baumgardt_2017}
{Sollima}, A., \& {Baumgardt}, H. 2017, \mnras, 471, 3668,
  \dodoi{10.1093/mnras/stx1856}

\bibitem[{{Terzi{\'c}} \& {Graham}(2005)}]{Terzic+2005}
{Terzi{\'c}}, B., \& {Graham}, A.~W. 2005, \mnras, 362, 197,
  \dodoi{10.1111/j.1365-2966.2005.09269.x}

\bibitem[{{Turk} \& {Lorimer}(2013)}]{Turk_Lorimer_2013}
{Turk}, P.~J., \& {Lorimer}, D.~R. 2013, \mnras, 436, 3720,
  \dodoi{10.1093/mnras/stt1850}

\bibitem[{Umbreit {et~al.}(2012)Umbreit, Fregeau, Chatterjee, \&
  Rasio}]{Umbreit_2012}
Umbreit, S., Fregeau, J.~M., Chatterjee, S., \& Rasio, F.~A. 2012, \apj, 750,
  31.
\newblock \url{http://dx.doi.org/10.1088/0004-637X/750/1/31}

\bibitem[{{VandenBerg} {et~al.}(2013){VandenBerg}, {Brogaard}, {Leaman}, \&
  {Casagrande}}]{VandenBerg+2013}
{VandenBerg}, D.~A., {Brogaard}, K., {Leaman}, R., \& {Casagrande}, L. 2013,
  \apj, 775, 134, \dodoi{10.1088/0004-637X/775/2/134}

\bibitem[{{Vasiliev} \& {Baumgardt}(2021)}]{Vasiliev+2021}
{Vasiliev}, E., \& {Baumgardt}, H. 2021, \mnras, 505, 5978,
  \dodoi{10.1093/mnras/stab1475}

\bibitem[{{Wijers} \& {van Paradijs}(1991)}]{Wijers+1991}
{Wijers}, R.~A.~M.~J., \& {van Paradijs}, J. 1991, \aap, 241, L37

\bibitem[{{Ye} {et~al.}(2019){Ye}, {Kremer}, {Chatterjee}, {Rodriguez}, \&
  {Rasio}}]{Ye_msp_2019}
{Ye}, C.~S., {Kremer}, K., {Chatterjee}, S., {Rodriguez}, C.~L., \& {Rasio},
  F.~A. 2019, \apj, 877, 122, \dodoi{10.3847/1538-4357/ab1b21}

\bibitem[{{Ye} {et~al.}(in preparation){Ye}, {Kremer}, {Ransom}, \&
  {Rasio}}]{Yeinprep}
{Ye}, C.~S., {Kremer}, K., {Ransom}, S.~M., \& {Rasio}, F.~A. in preparation

\bibitem[{{Ye} {et~al.}(2022){Ye}, {Kremer}, {Rodriguez}, {Rui}, {Weatherford},
  {Chatterjee}, {Rasio}, \& {Fragione}}]{Ye_47tuc_2021}
{Ye}, C.~S., {Kremer}, K., {Rodriguez}, C.~L., {et~al.} 2022, \apj, 931, 84,
  \dodoi{10.3847/1538-4357/ac5b0b}

\end{thebibliography}

\end{document}